\pdfoutput=1
\documentclass[onecolumn]{article}
\usepackage[utf8]{inputenc}
\usepackage{amsmath}
\usepackage{amssymb}
\usepackage[numbers]{natbib}
\usepackage[hidelinks]{hyperref}
\usepackage{geometry}
\usepackage{subcaption}
\usepackage{graphicx}
\usepackage{comment}
\usepackage{multirow}
\usepackage{algorithm}
\usepackage{algpseudocode}
\usepackage{xcolor}
\usepackage{soulutf8}
\newif\ifmarked
\markedfalse
\sethlcolor{yellow}
\soulregister{\cite}{7}
\soulregister{\citep}{7}
\soulregister{\citet}{7}
\soulregister{\ref}{7}
\soulregister{\eqref}{7}
\soulregister{\label}{7}
\soulregister{\url}{7}
\definecolor{revblue}{RGB}{0,60,170}
\newcommand{\rev}[1]{\ifmarked\hl{#1}\else#1\fi}
\newcommand{\revblock}[1]{\ifmarked\begingroup\color{revblue}#1\endgroup\else#1\fi}

\title{Time-Window Noise2Noise: Self-Supervised Blind Denoising of Mechanical Vibration and Impact Signals}

\author{
    \Large \textbf{Vinicius S. Vianna\textsuperscript{1}, Tiago H. Machado\textsuperscript{2} and Ilmar F. Santos\textsuperscript{3}} \\
    \vspace{0.2cm}
    \normalsize University of Campinas, Brazil\textsuperscript{1,2} and Technical University of Denmark, Denmark\textsuperscript{3} \\
    \small v178334@dac.unicamp.br\textsuperscript{1}, tiagomh@fem.unicamp.br\textsuperscript{2} and ilsa@dtu.dk\textsuperscript{3}
}

\date{}

\begin{document}

\maketitle
\begin{abstract}
Vibration and impact measurements in mechanical systems are invariably corrupted by noise, which degrades every quantity derived from them, such as modal parameters and contact forces. Classical filters rely on assumptions about the noise or the signal, and linear filters cannot remove the noise that shares the frequency band of the structural response. This work presents Time-Window Noise2Noise (TWN2N), a self-supervised method for blind denoising of signals produced by deterministic mechanical systems. The method requires neither a clean reference signal nor paired noisy observations, nor a measured noise variance or distribution: by reconstructing each time instant from its temporal neighborhood while masking the instant itself, a compact \rev{autoencoder-based network} is encouraged to exploit the coherent temporal and multichannel structure of the measurement rather than the noise. The formulation requires no model of the system and is assessed on two benchmarks with exactly known ground truth: the impulse response of a linear 3-degree-of-freedom system and a nonlinear Hunt--Crossley impact. Over repeated realizations, TWN2N exceeds a zero-phase low-pass filter at low and moderate input signal-to-noise ratio (SNR) and gives the largest gain inside the structural band, which a linear filter largely preserves, among the Savitzky--Golay and wavelet references; the advantage vanishes at high input SNR and under brown noise. The benefit propagates to identification: the damping ratio of the weakest mode, overestimated several-fold on the raw signals, is recovered close to its reference value after denoising.
\end{abstract}

\noindent \textbf{Keywords:} Blind denoising; Self-supervised learning; Mechanical vibration; Natural frequency and damping ratio identification; Contact dynamics; Autoencoder.

\section{Introduction}

The extraction of clean dynamic signals from noisy measurements is a recurring computational problem across structural dynamics, vibration analysis, and impact mechanics. In practice, the acquired signal is invariably corrupted by noise arising from sensor limitations, environmental interference, electronic disturbances, and analog-to-digital conversion. This corruption propagates to every quantity computed from the signal (natural frequencies, damping ratios, and contact forces) and can render their estimates unreliable or, at high noise levels, effectively unidentifiable. A robust, general-purpose denoising stage in the computational pipeline therefore directly improves the reliability of the downstream engineering analysis.

Several methods have been proposed to remove noise and recover the true underlying signal. A wide class of these approaches shares the same fundamental premise: denoising is achieved by averaging, frequently performed locally, as established by the classical Gaussian smoothing model \cite{lindenbaum1994gabor}. \rev{The classical methods of this class are digital filters. A digital filter is a discrete-time system that modifies the frequency content of a sampled signal; in its usual form it is linear and time-invariant, and it is then completely described by its frequency response \cite{oppenheim2010discrete}. The low-pass and Savitzky--Golay filters are of this kind, and wavelet thresholding applies a fixed nonlinear rule to the coefficients of a bank of such filters. In all of them, the operation applied to the signal is fixed in advance by a few design parameters.} However, traditional techniques, including wavelet-based methods \cite{donoho1994ideal} and kernel-based filters \cite{buades2005non,savitzky1964smoothing}, require assumptions about the noise statistics or the signal characteristics that may not hold in real-world applications: the VisuShrink threshold, for instance, requires the standard deviation of the noise and presumes a signal that is sparse in the wavelet basis; a low-pass filter requires a cutoff that must be known in advance and attenuates whatever noise and signal lie above and around it; and the Savitzky--Golay filter presumes that the signal is locally polynomial over its window, which must therefore be short against the fastest oscillation. Moreover, they typically struggle with extremely low signal-to-noise ratios (SNRs), nonstationary noise conditions, and colored noises, commonly encountered in dynamic testing. 

Data-driven, learning-based methods have become a prominent computational alternative to model-based filtering across engineering; for instance, Fan \textit{et al.}~\cite{fan2020vibration} employed residual convolutional networks to denoise vibration signals for structural health monitoring. Such supervised approaches, however, require large sets of clean reference signals for training, a requirement rarely met in practice, since the noise-free ground truth of an experimental signal is, in general, inaccessible.

The advent of deep learning has offered data-driven alternatives, initially pioneered by Denoising Autoencoders (DAEs) \cite{vincent2008extracting}.
In the specific context of image denoising, Jain \textit{et al.} \cite{jain2008natural}
were among the first to apply Convolutional Neural Networks (CNNs) to this task, establishing a setup where denoising is treated as a regression problem to minimize the loss between the prediction and clean ground truth data. Building on this, Zhang \textit{et al.} \cite{zhang2017beyond}
set a new performance standard with a very deep convolutional architecture built on residual learning. Their residual learning approach predicts the noise component directly, rather than the clean signal itself, enabling a single network to handle a wide range of noise intensities.

However, standard supervised models rely on clean ground truth signals, a requirement rarely satisfied in experimental mechanics. This constraint motivated the emergence of the blind denoising framework Noise2Noise (N2N) \cite{lehtinen2018noise2noise}, which learns from paired noisy observations. However, since acquiring multiple independent realizations of the same dynamic event is often prohibitive, research has pivoted toward Internal Statistics Methods\footnote{Internal Statistics Methods refer to a category of algorithms that do not require prior training on external ground truth data. Instead, they extract the required signal information directly from the internal statistical dependencies of the noisy test data itself.}. Although single-shot techniques within this category, such as Noise2Void \cite{krull2019noise2void} and Noise2Self \cite{batson2019noise2self}, successfully eliminate the need for paired data, they predominantly target spatial dependencies in static images, thereby neglecting the temporal coherence inherent to dynamical systems.

Blind-spot and Noise2Noise-type training has since been carried to one-dimensional measurements. van den Ende \textit{et al.}~\cite{vandenende2023self} removed spatially incoherent noise from distributed acoustic sensing records with a J-invariant self-supervised network, Lapins \textit{et al.}~\cite{lapins2024dasn2n} trained a Noise2Noise network on pairs of co-located channels of the same kind of data, and Li \textit{et al.}~\cite{li2025blind} combined self-supervised and transfer learning for the same purpose; Wang \textit{et al.}~\cite{wang2024seismic} applied self-supervised training to the denoising of seismic records; Langarica and N\'u\~nez~\cite{langarica2023contrastive} proposed a blind denoising autoencoder for industrial sensor streams; and Yang \textit{et al.}~\cite{yang2024blind} addressed the blind denoising of rotating-machinery vibration signals. These works confirm that self-supervision can replace clean targets for time series, but they treat each channel as a generic sequence: to our knowledge, none of them uses the deterministic evolution of a mechanical state, measured simultaneously on several channels, as the structure that separates the signal from the noise.

This work addresses this gap by carrying the blind-spot construction to deterministic dynamical systems. Time-Window Noise2Noise \rev{(TWN2N)} is a self-supervised \rev{autoencoder-based network} trained to reconstruct the state of a mechanical system at one instant from the neighboring observations of all its simultaneously acquired channels, with the instant itself masked, so that the network is encouraged to exploit the coherent temporal and multichannel structure of the measurement rather than the noise, without clean targets or paired data. \rev{TWN2N is not a digital filter in the sense defined above: its operation is a nonlinear mapping learned from the noisy measurement itself, and not fixed in advance by a few design parameters. It is therefore presented here as a denoising method, and the low-pass, Savitzky--Golay, and wavelet (VisuShrink) filters are the references against which it is compared.} Unlike spatial denoising methods that rely on statistical redundancy, the approach leverages the temporal structure of deterministic systems, enabling signal recovery and reliable estimation of natural frequencies and damping ratios at input SNRs between 5 and 20 dB, where linear filtering cannot separate the noise from the response inside the structural band. The contribution is therefore not a new network but the temporal, multichannel instance of the blind-spot principle: the neighborhood is a time window rather than a spatial patch, and one network reconstructs all the channels jointly instead of one network per channel, together with a characterization of where in the spectrum, for which noise types, and for which damping levels the gain is obtained.

The remainder of this paper is organized as follows: Section \ref{Methodology} details the Time-Window Noise2Noise methodology and the synthetic datasets used for validation. Section \ref{Results} presents the quantitative and statistical performance of the method across various noise levels. Finally, Section \ref{Conclusions} discusses implications and future perspectives.

\section{Method and dataset description}
\label{Methodology}

This section outlines the comprehensive methodology adopted in this study. First, the theoretical framework and the specific architecture of the proposed Time-Window Noise2Noise model are detailed, highlighting its distinct ability to exploit temporal correlations in dynamic signals compared to standard autoencoders. Subsequently, the generation of the synthetic datasets used for validation is described. These datasets, representing a linear multi-degree-of-freedom system and a nonlinear contact experiment, serve as ground truth baselines to rigorously evaluate the denoising performance under varying signal-to-noise ratios.

\subsection{Autoencoder-based architecture}
\label{sec:architecture}

An autoencoder is a self-supervised network that learns to reproduce its own input after forcing it through an information bottleneck \cite{goodfellow2016deep,majumdar2018blind}. It is built from an encoder $F_{en}$, which maps the input onto a compact latent code of reduced dimensionality, and a decoder $F_{de}$, which reconstructs the original signal from that code.

When such an architecture is employed for denoising, the conventional training scheme pairs a corrupted observation at the input with its uncorrupted counterpart as the regression target \cite{vincent2008extracting, chiang2019noise,  ashfahani2020devdan, cho2013simple}. Clean references of this kind are seldom available outside controlled environments. The stacked sparse denoising autoencoder (SSDA), for instance, attains its performance only after being exposed to large collections of clean samples \cite{cho2013simple}, and a distinct model has to be retrained whenever the noise statistics or the signal type change. In experimental mechanics, where uncorrupted reference signals essentially never exist, this supervised paradigm becomes impractical.

These limitations motivated the family of blind denoising strategies, which dispense with clean targets and require no prior knowledge of the noise type or intensity, making them considerably better suited to real measurements. Along this line, Majumdar \cite{majumdar2018blind} proposed a blind autoencoder that enforces sparsity in the latent representation through an $L1$-norm penalty, reporting improvements over denoising based on Singular Value Decomposition (SVD). More broadly, promoting sparsity in the latent features has repeatedly been found to benefit reconstruction quality \cite{cho2013simple}.

Unlike standard autoencoders that process each time instant independently, our proposed architecture exploits temporal context. By presenting sequential observations as input, the model captures the deterministic dependencies governing the system's evolution. This temporal window acts as a structural prior, guiding the network toward solutions consistent with the physics of the data. For autonomous dynamical systems, the time evolution of the clean state vector $X_0(t)$ is governed by a deterministic function $\mathcal{F}$:
\begin{equation}
    \dot{X}_0(t_i) = \dfrac{d}{dt}X_0(t_i) = \mathcal{F}(X_0(t_i))~.
\end{equation}

\begin{figure}[h!]
    \centering
    \includegraphics[width=0.86\columnwidth]{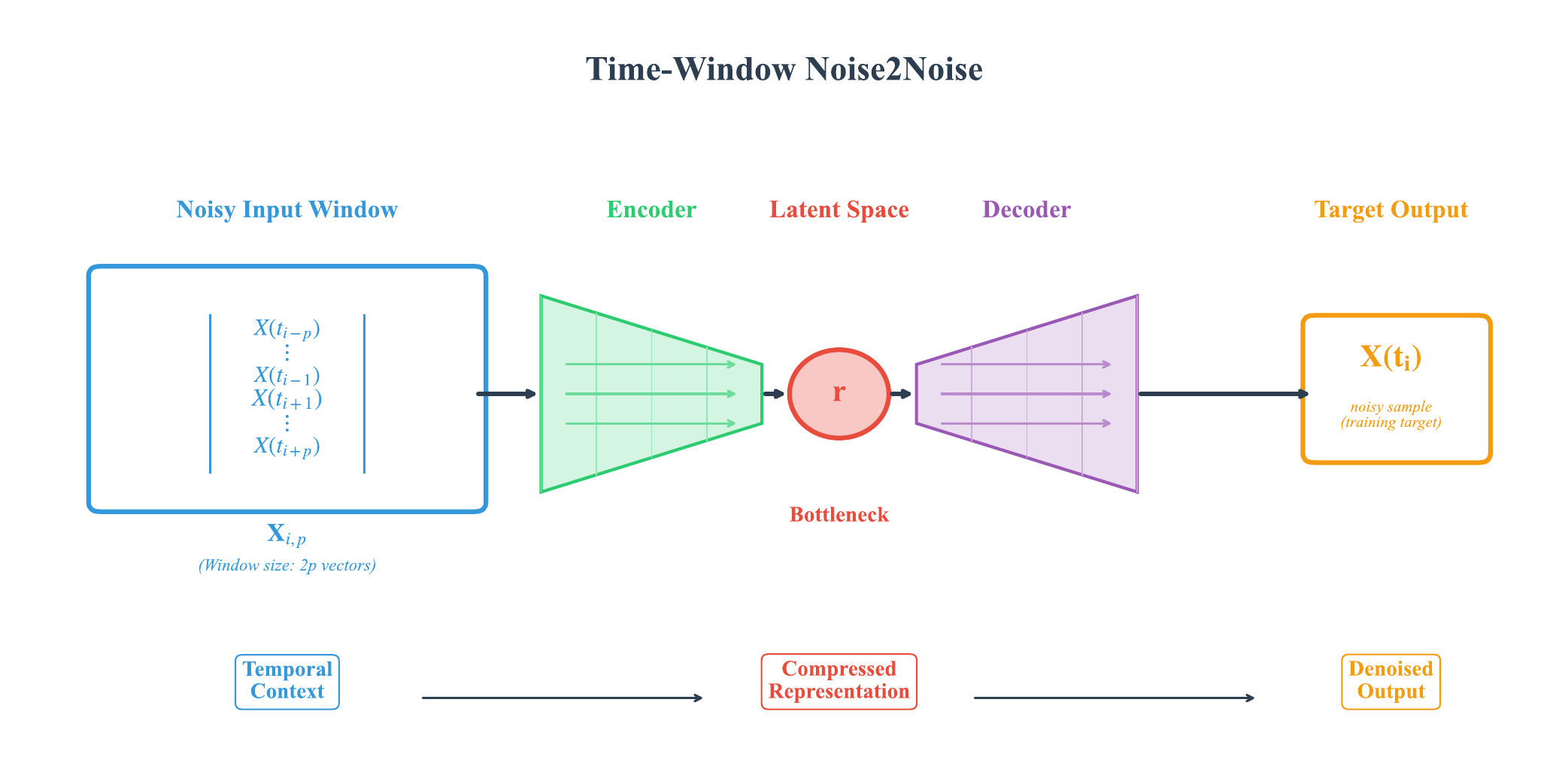}
    \caption{Schematic representation of the Time-Window Noise2Noise architecture.}
    \label{fig:autoencoder_architecture}
\end{figure}

The measured signal is the clean state corrupted by additive noise, $X(t) = X_0(t) + \varepsilon(t)$; the notation $X_0$ for the clean state, $X$ for the measured (noisy) sample, and $Y$ for the network output is used throughout. The diagram of the proposed Time-Window Noise2Noise architecture is depicted in Figure \ref{fig:autoencoder_architecture}. To estimate the state at time $t_i$ the network input is defined as a vector $\mathbf{X}_{i,p}$ that includes only the neighboring past and future measured samples, explicitly excluding the current sample $X(t_i)$:
\begin{equation}
    \mathbf{X}_{i,p} = [X(t_{i-p}), \dots, X(t_{i-1}), X(t_{i+1}), \dots, X(t_{i+p})]~,
    \label{Eq_p}
\end{equation}
where $p$ represents the window size (lag and lead steps).

This modeling approach is inspired by the Noise2Void framework \cite{krull2019noise2void} and its use of a "blind-spot" architecture. In Noise2Void, each pixel is estimated using only its spatial neighborhood while masking the central pixel itself, to prevent the network from learning a trivial identity mapping. Similarly, by explicitly excluding $X(t_i)$ from the input, and utilizing a bottleneck dimension significantly smaller than the dimension of a single input frame, we prevent the model from learning the identity function. Consequently, the network is encouraged to exploit the coherent temporal structure of the signal to predict $X(t_i)$. Distinct from image denoising, which exploits spatial redundancy, our method leverages the temporal determinism of physical dynamics. TWN2N therefore builds on Noise2Void but differs from it in three respects. First, the neighborhood is temporal: the $p$ past and $p$ future samples of the masked instant, instead of the spatial patch around a pixel. Second, the input and the output comprise all simultaneously acquired channels, so that a single network reconstructs the whole state vector at $t_i$ and can exploit the relations between channels (here the positions and accelerations of the three masses, which are linked by the equations of motion); a channel-wise blind-spot network would require one network per channel, each seeing only the past and future of its own channel. Third, a bottleneck smaller than the output dimension is imposed on the code, as a complementary constraint to the masking.

Since the physical signal $X_0(t)$ adheres to this smooth, deterministic evolution rule, while the stochastic noise $\varepsilon$ acts as an uncorrelated disturbance, the network effectively distinguishes between the two to predict the missing $X(t_i)$. The construction rests on one explicit condition and carries one limitation. With the $L_1$ loss used here, the population predictor targets the conditional median of the noisy target given the observed window. Independence of the noise at the masked instant from the noise in the window, with zero-median or symmetric noise, reduces the bias introduced by the noisy target, but equality with the conditional median of the clean sample requires additional symmetry or conditional-determinism assumptions and is not guaranteed in general. Recovery of the clean sample additionally requires the clean central state to be determined by the noisy context, which the noise in the neighboring samples prevents in part; the residual that remains at high input SNR (Section~\ref{sec:3dof_results}) is this reconstruction floor. The network contains no governing equation or physics-based loss: it exploits the temporal and multichannel coherence of the measured dynamics, and it is not physics-informed in the strict sense. Section~\ref{sec:noise_types} shows what happens when the independence condition is violated by temporally correlated noise.

The training target is the measured (noisy) sample at the instant $t_i$:
\begin{equation}
    X(t_i) = X_0(t_i) + \varepsilon(t_i)~,
\end{equation}
where $X_0(t_i)$ is the clean value and $\varepsilon(t_i)$ is the additive random noise. A key observation is that this target is the noisy sample itself, not the underlying clean value $X_0(t_i)$. As a consequence, driving the training loss to its absolute minimum does not coincide with optimal denoising; carried too far, the optimization would merely reproduce the noisy measurement. The gap between this noisy target and the desired clean output widens as the noise level grows, which is the reason for the early-stopping rule described at the end of this section and in Appendix~A.

The input vector $\mathbf{X}_{i,p}$ is flattened and compressed by the encoder into a latent code $r$:
\begin{equation}
    r = F_{en}(\mathbf{X}_{i,p})~.
\end{equation}

The latent dimension of the vector $r$ is deliberately chosen to be smaller than the output dimension (e.g. $\dim(r)=4$ for a 6-dimension output). As mentioned, this compressed representation favors the coherent part of the input over the high-dimensional noise.

The decoder then expands this latent code $r$ back to an estimate of the target instant:
\begin{equation}
    Y(t_i) = F_{de}(r)~.
\end{equation}

In the training phase, the noisy observation at the present instant $X(t_i)$ is the target. The objective is to optimize the weights $w$ of $F_{en}$ and $F_{de}$ to minimize the $L_1$ error norm between the predicted output and the noisy target:
\begin{equation}
    w = \operatorname*{argmin}_w \|Y(t_i) - X(t_i)\|_1~.
\end{equation}

The $L_2$ loss squares errors, thereby amplifying the impact of outliers and increasing noise sensitivity. In contrast, the $L_1$ norm has been shown to be more robust in previous studies \cite{goodfellow2016deep, majumdar2018blind,cho2013simple}. Being less sensitive to large residuals, the $L_1$ norm makes the model more resilient to noisy targets, allowing it to better tolerate discrepancies between noisy observations and the underlying clean signal.

Finally, to mitigate the challenge of learning from noisy targets, training duration is regulated by early stopping. Since neural networks tend to learn simpler patterns (such as the deterministic system dynamics) before fitting complex random data (noise), prolonged training would eventually lead the model to "memorize" the noise to further minimize the loss. At the population level the masked construction cannot reproduce noise that is independent of the window; with a finite record and an overparameterized network, however, prolonged training lets the model fit the particular noise realization of the training windows, a finite-sample effect that early stopping limits. In our experiments, this tendency to overfit the noise was the dominant failure mode, and it became markedly more pronounced as the noise level increased. Early stopping effectively prevents this overfitting, allowing the model to capture the deterministic signal structure without learning the noise distribution.

\subsection{Algorithm and software implementation}
\label{sec:software}
The complete procedure is summarized in Algorithm~\ref{alg:twn2n}. The method is deliberately simple in architecture and framework-agnostic: it requires only a \rev{feed-forward autoencoder-based network} and a lagged (time-windowed) construction of the training set, and it is therefore straightforward to reproduce and to embed in existing computational pipelines.

\begin{algorithm}[h!]
\caption{Time-Window Noise2Noise (TWN2N).}
\label{alg:twn2n}
\begin{algorithmic}[1]
\Require noisy signal $\{X(t_i)\}_{i=1}^{n}$; window size $p$; latent dimension \rev{$\dim(r)$}
\Ensure denoised signal $\{\hat{X}(t_i)\}_{i=1}^{n}$
\For{each instant $t_i$}
  \State assemble the blind-spot input $\mathbf{X}_{i,p}=[X(t_{i-p}),\dots,X(t_{i-1}),X(t_{i+1}),\dots,X(t_{i+p})]$ \Comment{\rev{$X(t_i)$} masked}
\EndFor
\State standardize inputs $\mathbf{X}_{i,p}$ and targets $X(t_i)$
\State initialize encoder $F_{en}$ and decoder $F_{de}$ with bottleneck \rev{$\dim(r)$}${}<\dim(\text{output})$
\While{early-stopping criterion not met}
  \State update weights to minimize $\sum_i \lVert F_{de}(F_{en}(\mathbf{X}_{i,p})) - X(t_i)\rVert_1$ \Comment{Adam}
  \State monitor validation loss; keep best weights
\EndWhile
\State $\hat{X}(t_i) \gets F_{de}(F_{en}(\mathbf{X}_{i,p}))$ for all $i$ \Comment{single feed-forward pass}
\end{algorithmic}
\end{algorithm}

The reference implementation\rev{~\cite{vianna2026twn2ncode}} uses the Keras/TensorFlow framework; the encoder--decoder layer configuration and all training hyperparameters are reported in Appendix~A. The TWN2N network used for the 3-DOF benchmark has $17{,}218$ trainable parameters; training is performed once per \rev{realization, jointly on its three impulse-response records,} and, over the 450 trainings of the 3-DOF study (records of 1 to 8 s at 1024 Hz, two central processing unit (CPU) cores, no graphics processing unit (GPU)), took between 0.6 and 10 minutes, with a median of 2.4 minutes (Appendix~A). Inference is one forward pass of the network per time sample, so its cost grows in proportion to the number of samples: denoising the three records of one realization (3,072 samples each, 9,216 in total, six channels) takes 0.35 s on a desktop CPU, against 2 to 13 ms for the Savitzky--Golay, low-pass, and wavelet filters of \rev{Sections~\ref{sec:3dof_results} and}~\ref{sec:classical}. The network is thus slower than a classical filter at inference by two orders of magnitude, but this cost is negligible against the training time, and the method requires no \rev{prior knowledge of the natural frequencies or of the noise level, which the classical filters need to set their cutoff, window, or threshold}, its own design choices ($p$, \rev{$\dim(r)$}, architecture, patience) being fixed a priori.

\subsection{Datasets}
The proposed method was evaluated on two synthetic datasets: a linear 3-DOF mass-spring-damper system and a nonlinear impact system. Subsequently, noise was added to simulate experimental conditions: white Gaussian noise to both datasets, and pink, brown, and quantization noise to the 3-DOF dataset (Section~\ref{sec:noise_types}). The noisy signal $X$ was generated from the clean reference $X_0$ as follows:
\begin{equation}
    X = X_0 + \sqrt{\frac{\sigma^2(X_0)}{SNR}} \cdot \mathcal{N}(0,1)~,
\end{equation}
where $\sigma^2(\cdot)$ denotes the variance operator, and $\mathcal{N}(0,1)$ represents a standard Gaussian noise vector with the same dimensions as the input data $X$. The Signal-to-Noise Ratio (SNR) determines the intensity of the corruption and is defined as:
\begin{equation}
    SNR = \frac{\sigma^2(X_0)}{\sigma^2(\varepsilon)}~,
\end{equation}
where $\varepsilon$ represents the noise component ($X - X_0$). All SNR values in this work are this ratio expressed in decibels, $\mathrm{SNR}_{\mathrm{dB}} = 10\log_{10}\left[\sigma^2(X_0)/\sigma^2(\varepsilon)\right]$, and the output SNR of a processed signal $Y$ is computed with the same expression with $\varepsilon$ replaced by the residual $Y - X_0$. To evaluate the robustness of the Time-Window Noise2Noise model, varying SNR levels were introduced to both datasets. High SNR conditions ($> 25$ dB) are typical of controlled laboratory tests such as impulse response characterization. Conversely, cases with lower SNR ($< 15$ dB) are typical of measurements in challenging contact/impact tests, where high-frequency structural responses are frequently obscured by sensor limitations and nonlinear interactions.

The use of synthetic benchmarks is a deliberate methodological choice, standard in computational engineering: an exactly known ground truth with precisely controlled noise levels is required to quantify denoising performance unambiguously and reproducibly, isolating the behavior of the method from modeling assumptions and experimental confounders. Validating novel data-driven architectures first on such controlled benchmarks is established practice \cite{lehtinen2018noise2noise, krull2019noise2void, batson2019noise2self, majumdar2018blind}. The validation of the method on laboratory data (impact-hammer tests and operational modal analysis) is not part of this manuscript; the claims made here are restricted to the two numerical benchmarks.

\subsubsection{Linear 3-DOF mass-spring-damper system}
\label{sec:3dof_dataset}

The first dataset comprises the impulse response of a linear 3-degree-of-freedom (3-DOF) mass-spring-damper system. The governing equation of motion is given by:
\begin{equation}
    \mathbf{M}\ddot{\mathbf{x}}(t) + \mathbf{C}\dot{\mathbf{x}}(t) + \mathbf{K}\mathbf{x}(t) = \mathbf{f}(t)~.
\end{equation}

The system matrices (Mass $\mathbf{M}$, Stiffness $\mathbf{K}$, and Damping $\mathbf{C}$) are defined based on the physical connectivity of the masses, springs, and dampers. The mass matrix is diagonal:
\begin{equation}
    \mathbf{M} = \begin{bmatrix} 
    m_1 & 0 & 0 \\ 
    0 & m_2 & 0 \\ 
    0 & 0 & m_3 
    \end{bmatrix}~.
\end{equation}

\begin{figure}[h!]
    \centering
    \begin{subfigure}[c]{0.28\textwidth}
        \centering
        \includegraphics[width=\linewidth, keepaspectratio]{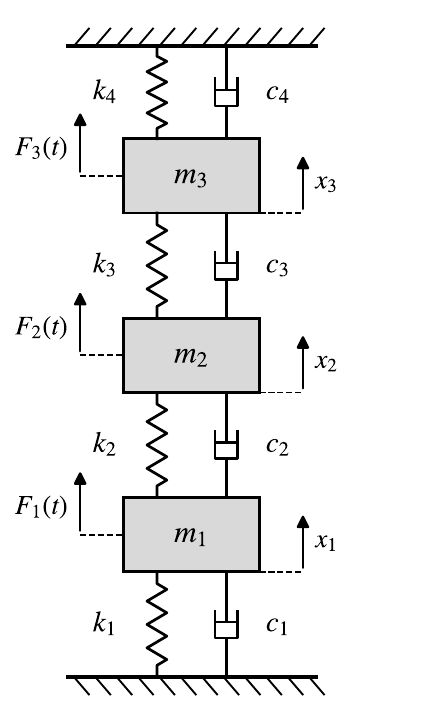}
        \caption{Schematic representation of the discrete 3-DOF system.}
        \label{fig:3dof_system}
    \end{subfigure}
    \hfill 
    \begin{subfigure}[c]{0.65\textwidth}
        \centering
        \includegraphics[width=\linewidth, keepaspectratio]{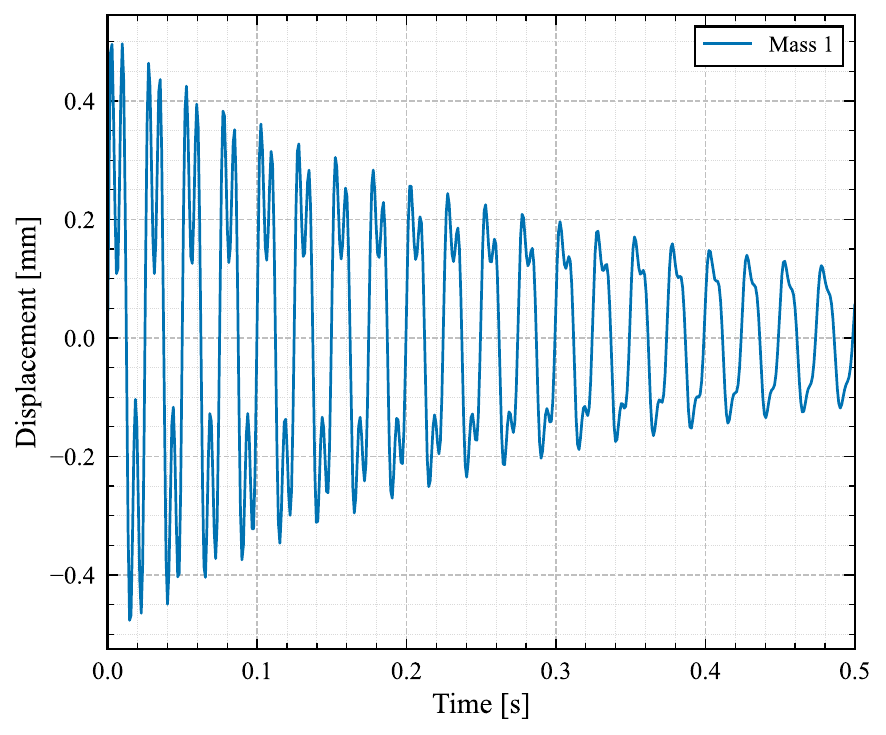}
        \caption{Time-domain response of mass $m_1$ subject to impulse excitation (first 0.5 s of the record).}
        \label{fig:impulse_response}
    \end{subfigure}
    
    \caption{Simulated 3-DOF system: (a) Mechanical model and (b) its corresponding dynamic unit impulse response.}
    \label{fig:combined_view}
\end{figure}

The stiffness matrix $\mathbf{K}$ follows from the chain connectivity of Figure~\ref{fig:3dof_system}:
\begin{equation}
    \mathbf{K} = \begin{bmatrix} 
    k_1 + k_2 & -k_2 & 0 \\ 
    -k_2 & k_2 + k_3 & -k_3 \\ 
    0 & -k_3 & k_3 + k_4 
    \end{bmatrix}~.
\end{equation}

The damping matrix proposed is proportional (Rayleigh) damping:
\begin{equation}
    \mathbf{C} = \alpha\,\mathbf{M} + \beta\,\mathbf{K}~,
    \label{eq:rayleigh}
\end{equation}
so that the modal damping ratios are $\zeta_r = \alpha/(2\omega_r) + \beta\,\omega_r/2$; $\alpha$ and $\beta$ are set from prescribed damping ratios at the first and third modes. The discrete dampers $c_1$--$c_4$ drawn in Figure~\ref{fig:3dof_system} are therefore illustrative of the dissipation mechanism rather than a literal representation of $\mathbf{C}$. Section~\ref{sec:damping_ablation} varies $(\alpha,\beta)$ over the range of damping ratios found in metallic structures.

For the numerical simulation and data generation the following physical parameters were adopted: masses $m_1 = m_2 = m_3 = 3.0$ kg; stiffness coefficients $k_1 = 302$, $k_2 = 1067$, $k_3 = 2998$, and $k_4 = 284$ kN/m; and Rayleigh coefficients $\alpha = 4.322$ s$^{-1}$ and $\beta = 11.37\times10^{-6}$ s, which give damping ratios $\zeta_1 = \zeta_3 = 0.010$ and $\zeta_2 = 0.0072$, i.e. the lightly damped regime of welded or jointed metallic structures. The natural frequencies are $f_1 = 40.1$, $f_2 = 120.1$, and $f_3 = 239.9$ Hz ($252.3$, $754.6$, and $1507.4$ rad/s). The responses are sampled at $f_s = 1024$ Hz, so the highest mode is sampled 4.3 times per period; this sparse sampling does not hinder the blind-spot reconstruction, because for a linear system sampled at a fixed interval the measured channels at the neighboring instants determine the state at the masked instant exactly (the state transition over one sampling interval is the matrix exponential of the system matrix), so the finite-difference argument of Section~\ref{sec:3dof_results} fixes the minimum window and not an oversampling requirement. Each record lasts $T = 3$ s, about seven time constants of the slowest mode ($1/(\zeta_1\omega_1) = 0.40$ s), so that the record covers the free decay without a long noise-only tail. These values were chosen so that the structural response occupies a substantial fraction of the measured bandwidth ($f_3 \approx 0.47\,f_s/2$): a linear low-pass filter can then remove only the noise above $f_3$, and any improvement inside the structural band must come from the dynamics learned by the network (Section~\ref{sec:lowpass}).

The theoretical natural frequencies $\omega_n$ and damping ratios $\zeta$ were calculated analytically by solving the eigenvalue problem associated with the state-space representation of the system; these values serve as the ground truth for the identification study of Section~\ref{sec:modal_id}.

The dataset was generated by computing the unit impulse response function (IRF). The system is formulated in the state-space form:
\begin{equation}
    \dot{\mathbf{z}}(t) = \mathbf{A}\mathbf{z}(t) + \mathbf{B}u(t)~,
\end{equation}
where the state vector is $\mathbf{z}(t) = \{\mathbf{x}(t), \dot{\mathbf{x}}(t)\}^T$, and the system matrices $\mathbf{A}$ and $\mathbf{B}$ are constructed using $\mathbf{M}$, $\mathbf{C}$, and $\mathbf{K}$. The impulse response was obtained by setting  $u(t) = \delta(t)$, where $\delta(t)$ is the Dirac delta function\rev{, applied to each mass in turn}.

Figure \ref{fig:combined_view} shows the system schematic and its time-domain response to a unit impulse applied at $m_1$.

The system response was simulated, and varying levels of noise were superimposed on the output signals as described previously. This dataset specifically aims to test the method's ability to preserve the phase and amplitude relationships required for the identification of natural frequencies and damping ratios.

\subsubsection{Nonlinear impact experiment (Hunt and Crossley model)}
\label{sec:impact_dataset}

The second dataset simulates an impact experiment. The contact force was modeled using the classic Hunt and Crossley model \cite{hunt1975coefficient}, which accounts for the energy dissipation during impact through a nonlinear damping term.

The governing equation for the contact force $F_N$ is given by:
\begin{equation}
    F_N = K\delta^{3/2} + \chi\delta^{3/2}\dot{\delta}~.
    \label{eq:HC_force}
\end{equation}

\begin{figure}[h!]
    \centering
    \includegraphics[width=0.8\columnwidth]{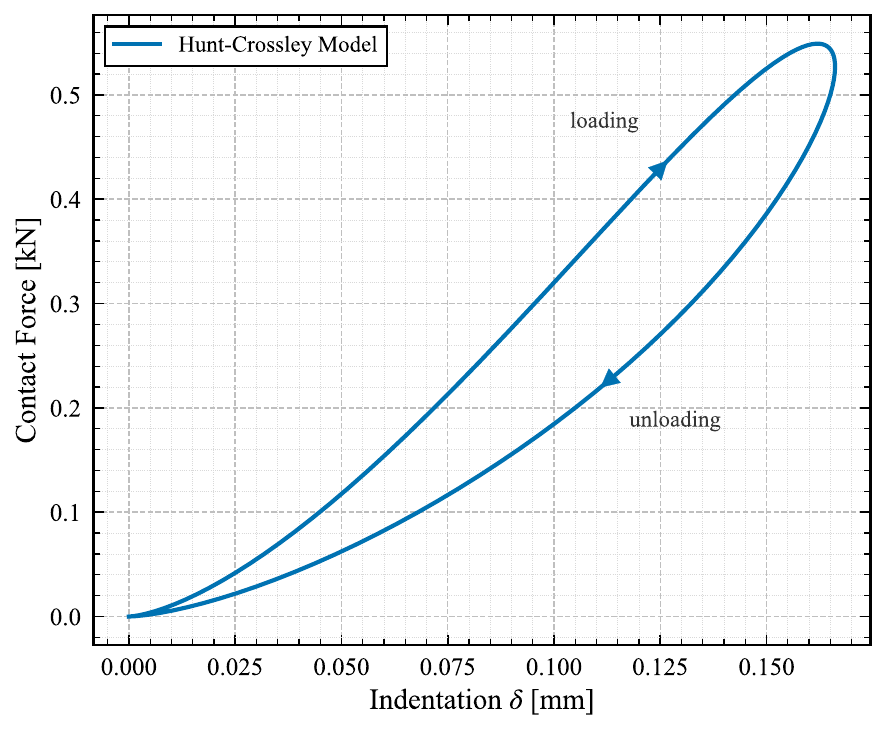}
    \caption{Characteristic force-indentation hysteresis loops generated by the Hunt and Crossley model.}
    \label{fig:histeretic}
\end{figure}

The hysteretic coefficient $\chi$ is defined to ensure consistency with the energy loss during impact:
\begin{equation}
    \chi = \frac{3}{2}(1-e) \frac{K}{v_{\text{approach}}}~,
    \label{eq:HC_chi}
\end{equation}
where $K$ is the stiffness, $\delta$ is the indentation depth, $\dot{\delta}$ is the indentation velocity, and $\chi$ is dependent on the coefficient of restitution $e$ and the initial impact velocity $v_{\text{approach}}$; $e$ is the nominal parameter of the law, and the rebound ratio of the simulated impact differs slightly from it. Figure \ref{fig:histeretic} illustrates the resulting nonlinear force-displacement hysteresis for the Hunt and Crossley model ($K = 2.46\times10^{8}$ N/m$^{1.5}$, $e = 0.76$, $v_{\text{approach}} = 0.4$ m/s). The enclosed area represents dissipated energy.

The dataset comprises displacement, velocity, and force time-histories generated for various initial impact velocities and restitution coefficients based on parameters from real experiments \cite{zhang2009validation}. Gaussian noise was added to these synthetic signals to emulate realistic sensor limitations.

\subsection{Numerical robustness evaluation}
\label{Uncertainty}
Because the two benchmarks provide an exactly known ground truth, the reliability of the method can be assessed directly and reproducibly, free of the confounders that hinder such quantification with real acquisitions. To this end, each noise condition is realized $N=30$ times \rev{in the 3-DOF benchmark and three times in the impact benchmark,} with independent noise samples and independent network initializations, and the quantities of interest are aggregated over the ensemble. This repeated-realization procedure is a Monte~Carlo propagation of the input-noise distribution to the output quantities. For a given quantity $q$, the tables report the ensemble mean $\bar q$ and the sample standard deviation $s(q)$ over the $N$ realizations, which is the run-to-run dispersion of a single estimate. Statements about the ensemble mean use its Type A standard uncertainty $u(\bar q) = s(q)/\sqrt{N}$ \cite{jcgm2008gum} and the 95\% confidence interval $\bar q \pm t_{0.975,N-1}\,u(\bar q)$, with $t_{0.975,29} = 2.05$ for $N = 30$, so that the half-width of the interval is $0.37\,s(q)$. Denoising performance is thus expressed both as a signal-to-noise-ratio gain and as the run-to-run statistical dispersion of the signal and of the quantities derived from it.

\section{Results}
\label{Results}

This section presents the evaluation of the proposed Time-Window Noise2Noise (TWN2N) architecture on the two case studies, the linear 3-DOF system and the nonlinear impact system. For the 3-DOF system, a standard autoencoder without temporal window is evaluated first as a baseline; the SNR gain of TWN2N is then quantified over the input noise levels and compared with that of a zero-phase low-pass filter \rev{with a fixed cutoff}, both inside and outside the structural band; the robustness to pink, brown, and quantization noise follows (Section~\ref{sec:noise_types}); the effect of denoising on the identification of natural frequencies and damping ratios with a classical time-domain estimator is assessed (Section~\ref{sec:modal_id}); the sensitivity to the damping level is examined (Section~\ref{sec:damping_ablation}); and the method is compared with Savitzky--Golay and wavelet filters (Section~\ref{sec:classical}). Section~\ref{sec:impact} reports the impact experiment.

\subsection{Noise suppression on the linear 3-DOF dataset}
\label{sec:3dof_results}

A standard autoencoder without temporal windowing was first evaluated as a baseline on the 3-DOF dataset. The input layer processed the instantaneous positions and accelerations of the three masses at time $t_i$, mapping them to an output of the same dimension, with a bottleneck (latent space) size of 4. Apart from the absence of the temporal window (6 instead of 24 input features), \rev{the network has the same depth, bottleneck, and activations as TWN2N} (Appendix A, 15,360 against 17,218 parameters) and the training protocol is the same. When trained with noisy impulse response signals (SNR of 20 dB) and optimized using the ADAM algorithm, the model yielded an output SNR of $22.06 \pm 0.95$ dB over thirty trials, a gain of 2.1 dB that stems from the redundancy among the six simultaneously acquired channels rather than from the dynamics. This result suggests that without temporal context, the network has no access to the evolution of the state, and its gain remains below that of a linear low-pass filter (\rev{2.9 dB} at this level, Table~\ref{tab:snr_lowpass}) and below the 3.1 dB of the windowed model presented next, motivating the usefulness of the proposed window-based approach.

The Time-Window Noise2Noise architecture was then evaluated. The temporal window size $p = 2$ was selected based on a physically grounded criterion rooted in the kinematics of mechanical systems. For a discrete-time signal with sampling interval $h$, the central-difference approximation of the instantaneous acceleration requires three consecutive position points:
\begin{equation}
    \ddot{x}(t_i) \approx \frac{x(t_{i-1}) - 2x(t_i) + x(t_{i+1})}{h^2}~.
\end{equation}

Since the central frame $x(t_i)$ is deliberately masked in the TWN2N input to prevent identity mapping, the network must infer it from neighboring observations. With $p = 2$, the input spans the four samples $x(t_{i \pm 1})$ and $x(t_{i \pm 2})$, providing the minimal symmetric context from which $x(t_i)$ can be reconstructed through the implicit physics of the differential equation governing the system. Setting $p = 1$ would reduce the input to only two neighboring frames, insufficient to capture the second-order dynamics encoded in the equations of motion. Thus, $p = 2$ is the smallest symmetric window that contains the three-point central-difference stencil of a second-order system once the central sample is masked; it is a physically motivated design choice, not a claim of optimality or universality. The window is therefore fixed at $p = 2$ in every experiment of this work, including the nonlinear impact dataset, and no search over $p$ was carried out, deliberately: selecting $p$ by maximizing the output SNR would require the clean signal, which is precisely what is unavailable in a real measurement, so any such tuning would \rev{rely on information that a practitioner does not have}. The choice rests instead on the finite-difference argument above, which holds for any second-order mechanical system regardless of its parameters.

The network input comprised these neighboring steps, yielding 24 features (2 state variables $\times$ 3 masses $\times$ 4 time points), which were encoded into a latent vector of dimension \rev{$\dim(r)$}${} = 4$ before reconstruction into a 6-dimensional output frame. This bottleneck dimension follows the usual autoencoder design rule: the latent dimension is kept smaller than the output dimension (\rev{$\dim(r)$}${} < 6$) to force compression, so that the decoder cannot reproduce its target through a near-identity map of a full-rank code and must instead encode the state of the system \cite{vincent2008extracting,goodfellow2016deep,majumdar2018blind}. The masked input already excludes a direct input-output identity, so the bottleneck is a complementary constraint rather than the only one. By setting \rev{$\dim(r)$}${} = 4$, a deliberate compression is introduced relative to the output space, encouraging the encoder to prioritize the coherent dynamical features over the high-dimensional stochastic noise component, which, when it is temporally uncorrelated, cannot be compressed by an architecture that exploits temporal coherence. Likewise, \rev{$\dim(r)$} was not tuned: the only requirement adopted is \rev{$\dim(r)$}${} < \dim(\text{output})$, so that the code must compress the state of the system, and \rev{$\dim(r)$}${} = 4$ for the 3-DOF system (\rev{$\dim(r)$}${} = 3$ for the impact dataset, whose output has five components) satisfies it. The claim of this work is thus that the method works with a window and a bottleneck chosen from these two principles alone, the finite-difference argument for $p$ and the bottleneck argument for \rev{$\dim(r)$}, and not that the specific values are optimal; the results that follow, obtained with the same $p$ in every experiment, on two datasets and over the range of damping levels of Section~\ref{sec:damping_ablation}, support this claim. Further details regarding the specific network architecture and training hyperparameters are provided in the Appendix.

\begin{figure}[h!]
    \centering
    \includegraphics[width=0.95\columnwidth]{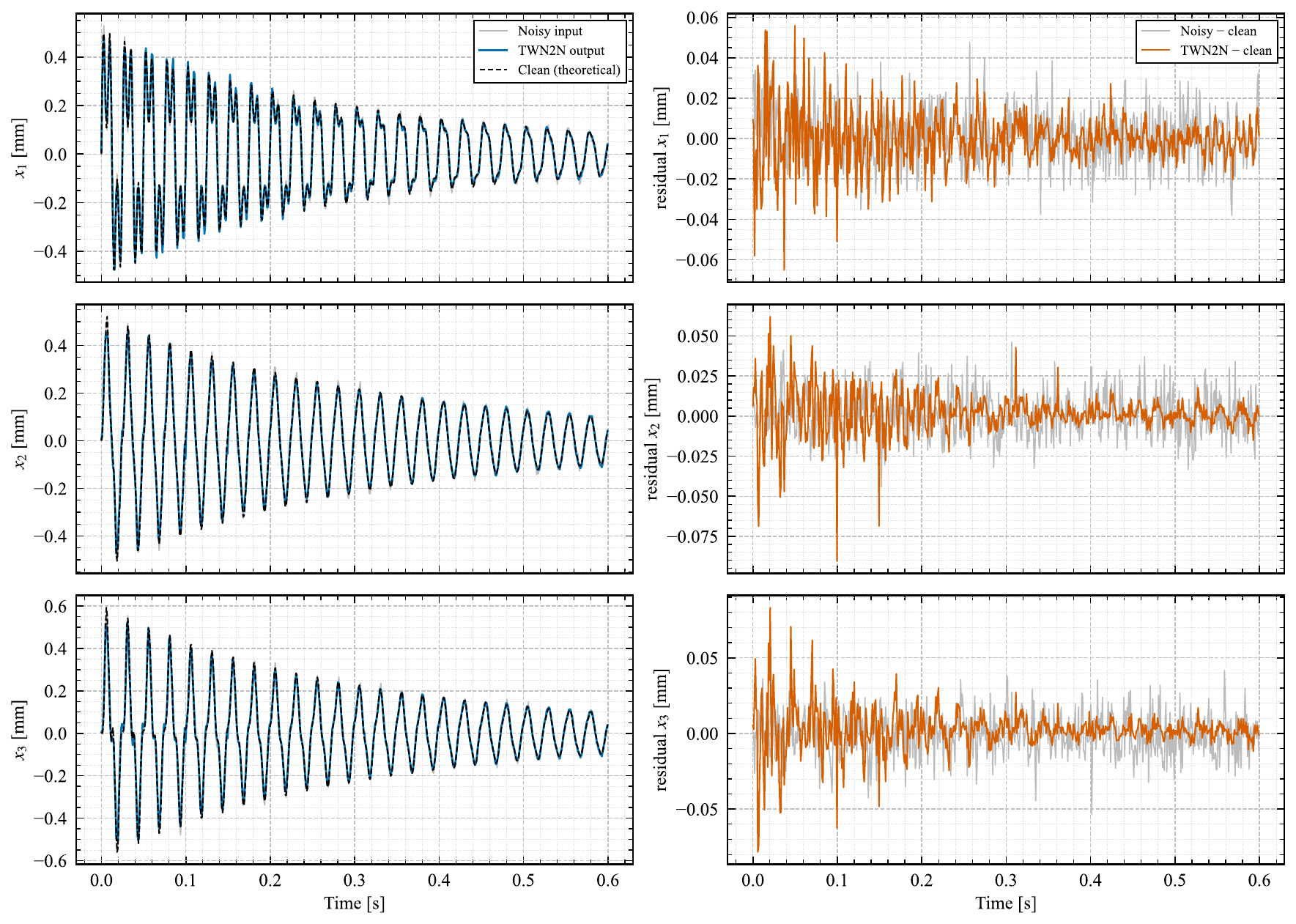}
    \caption{Per-mass comparison at 15 dB input SNR (impulse at $m_1$, first 0.6 s of the 3 s record): left, theoretical clean response overlaid with the noisy input and the TWN2N output; right, residuals with respect to the clean signal before and after denoising.}
    \label{fig:overlay_residuals}
\end{figure}

The qualitative performance of the method on the 3-DOF system can be assessed by comparing the ground truth with the input and the output of the network; Figure~\ref{fig:overlay_residuals} shows this comparison mass by mass for one realization at 15 dB input SNR, the level used for the identification study of Section~\ref{sec:modal_id}.

Visual inspection shows that the model output follows the clean response, and Figure~\ref{fig:overlay_residuals} makes the agreement explicit mass by mass at 15 dB input. In the window shown (first 0.6 s), the mean absolute residual of the three masses falls from 11.3, 11.5, and 11.5 to 9.6, 8.9, and 8.8 $\mu$m after denoising; over the full 3 s record it falls from 11.4, 11.2, and 11.3 to 4.4, 4.4, and 4.4 $\mu$m, i.e.\ the residual is reduced by a factor of 2.6 on average, more in the tail, where the signal is small, than in the first cycles. Over most of the decay the residual of the denoised signal is smaller than the noise, while a residual proportional to the signal remains during the first cycles, where the amplitude is largest; this is the reconstruction floor of the network, which bounds its performance at high input SNR (at 15 dB input the gain between the resonances is 7.5 to 10.2 dB, Table~\ref{tab:bands}). To quantify the performance, the model was evaluated across input SNR levels from 5 dB to 25 dB. For each level, 30 independent experiments were conducted, each with a new noise realization, a new random weight initialization and a new mini-batch order. The network is trained on the very noisy \rev{records} that it denoises: 90\% of the samples drive the gradient updates and the remaining 10\%, drawn at random, are held out only to monitor the validation loss that controls early stopping (Appendix~A); there is no separate test set, since the method is self-supervised and the clean signal is used solely to score the output. The 90/10 division is an in-record monitoring subset used only for early stopping: adjacent windows overlap, so it is not an independent generalization test, and the standardization is fitted on the whole record.

Table~\ref{tab:snr_comparison} summarizes the consolidated results, presenting the mean values and standard deviations of the output SNR and the resulting SNR gain. From 20 dB down to 5 dB input the gain is positive in every realization \rev{and increases from 3.1 dB at 20 dB to 8.9 dB at 5 dB; its mean is more than 20 times its standard uncertainty}. At 25 dB input the gain is negative \revblock{($-1.0$ dB, with a positive gain in only 6 of the 30 realizations)}: the noise is already below the reconstruction floor of the network, and the output is limited by the reconstruction error rather than by the noise. The operating range of the method on this system is therefore bounded above by an input SNR of about 20 dB. The standard deviations (0.6--1.0 dB) characterize the run-to-run dispersion of the output SNR over the ensemble of realizations, i.e. the joint effect of the noise realization and of the training randomness; the standard uncertainty of each mean gain ($s/\sqrt{N}$) is 0.1--0.2 dB.

\begin{table}[h!]
\centering
\caption{Statistical performance of the Time-Window Noise2Noise model (30 independent trials per noise level).}
\label{tab:snr_comparison}
\resizebox{0.75\columnwidth}{!}{%
\begin{tabular}{ccccccc}
\hline
\textbf{Input SNR} & \textbf{Mean Output} & \textbf{Std Dev} & \textbf{Mean Gain} & \textbf{Median Gain} & \textbf{Min Gain} & \textbf{Max Gain} \\
\textbf{(dB)} & \textbf{(dB)} & \textbf{(dB)} & \textbf{(dB)} & \textbf{(dB)} & \textbf{(dB)} & \textbf{(dB)} \\ \hline
25 & 24.04 & 1.04 & -0.96 & -0.98 & -3.33 & 1.19 \\
20 & 23.11 & 0.78 & 3.11 & 3.12 & 1.58 & 4.91 \\
15 & 21.01 & 0.68 & 6.01 & 6.02 & 4.21 & 7.12 \\
12 & 19.54 & 0.57 & 7.54 & 7.58 & 5.81 & 8.50 \\
10 & 18.09 & 0.69 & 8.10 & 8.24 & 6.17 & 8.91 \\
5 & 13.87 & 0.64 & 8.88 & 8.98 & 7.68 & 10.26 \\
\hline
\end{tabular}%
}
\end{table}

\begin{table}[h!]
\centering
\caption{TWN2N against the zero-phase Butterworth low-pass filter \rev{with a fixed cutoff at $1.05 f_3$} (LP) on the same 30 realizations per level: output SNR and gain of both methods in the full band, and gains of both methods inside the structural band $0$--$1.25 f_3$\rev{, most of which lies below the cutoff}. The better of the two methods is in bold.}
\label{tab:snr_lowpass}
\resizebox{0.8\columnwidth}{!}{%
\begin{tabular}{ccccccc}
\hline
\textbf{Input SNR} & \textbf{TWN2N output} & \textbf{TWN2N gain} & \textbf{LP output} & \textbf{LP gain} & \textbf{In-band gain} & \textbf{In-band gain} \\
\textbf{(dB)} & \textbf{(dB)} & \textbf{(dB)} & \textbf{(dB)} & \textbf{(dB)} & \textbf{TWN2N (dB)} & \textbf{LP (dB)} \\ \hline
25 & 24.04 & -0.96 & \revblock{\textbf{27.00}} & \revblock{\textbf{2.00}} & -0.14 & \revblock{\textbf{0.68}} \\
20 & \textbf{23.11} & \textbf{3.11} & \revblock{22.92} & \revblock{2.92} & \textbf{3.28} & \revblock{0.97} \\
15 & \textbf{21.01} & \textbf{6.01} & 18.25 & 3.25 & \textbf{5.67} & \revblock{1.06} \\
12 & \textbf{19.54} & \textbf{7.54} & 15.35 & 3.35 & \textbf{6.80} & \revblock{1.09} \\
10 & \textbf{18.09} & \textbf{8.10} & 13.39 & 3.39 & \textbf{7.40} & \revblock{1.09} \\
5 & \textbf{13.87} & \textbf{8.88} & 8.41 & 3.41 & \textbf{8.16} & \revblock{1.10} \\
\hline
\end{tabular}%
}
\end{table}

\begin{figure}[h!]
    \centering
    \includegraphics[width=0.95\columnwidth]{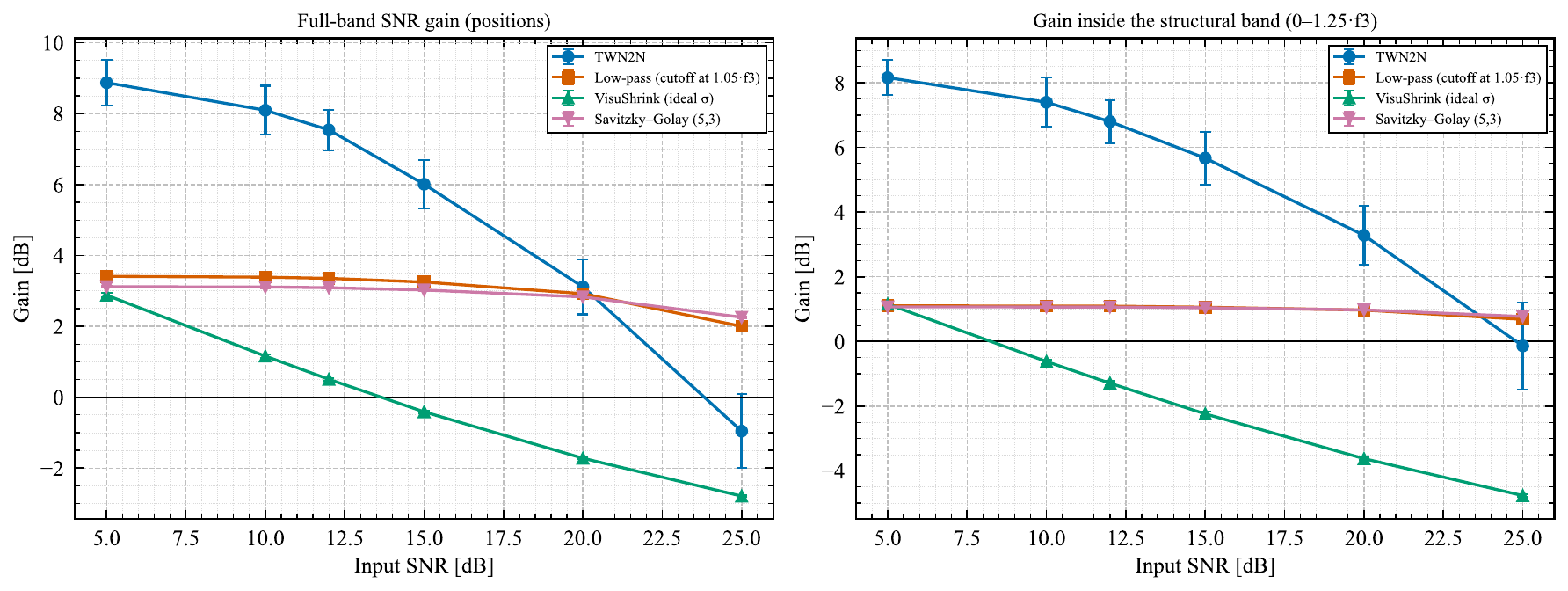}
    \caption{SNR gain versus input SNR for TWN2N, the low-pass filter \rev{with cutoff at $1.05 f_3$}, VisuShrink with the exact noise level and the Savitzky--Golay filter (mean $\pm$ standard deviation over 30 realizations): left, full band; right, inside the structural band $0$--$1.25 f_3$.}
    \label{fig:gain_methods}
\end{figure}

\label{sec:lowpass}
A linear low-pass filter is the natural benchmark for any denoiser of band-limited structural responses: with white noise, a zero-phase Butterworth filter (fourth order, applied forward and backward) with cutoff $f_c$ removes approximately the fraction $1 - f_c/(f_s/2)$ of the noise power (the brick-wall approximation of its response). \revblock{The cutoff is fixed at $f_c = 1.05 f_3$ in every experiment on the 3-DOF benchmark, under the assumption that the highest natural frequency of the structure is known in advance, so that the same filter is applied to all noise levels, noise types, and damping configurations. This cutoff lies just above the third resonance, where the zero-phase response already attenuates the signal by 3.8 dB. Among cutoffs from $1.05 f_3$ to $2 f_3$, it gives the filter its largest full-band gain from 15 dB input down; with the cutoff at $1.25 f_3$, for which the response is flat at $f_3$, the filter gains 2.6 dB at every level.} Table~\ref{tab:snr_lowpass} reports, for the same noisy realizations, the output SNR of the low-pass filter, together with the gains obtained inside the structural band $0$--$1.25 f_3$\rev{, most of which lies below the cutoff}. \rev{The low-pass filter gains 2.9--3.4 dB between 20 and 5 dB input, a value set by the fraction of the bandwidth above the cutoff, and 2.0 dB at 25 dB, where the attenuation of the third mode by the filter is no longer small compared with the noise.} TWN2N exceeds it by 2.8 dB at 15 dB input, 4.7 dB at 10 dB and 5.5 dB at 5 dB, \rev{exceeds it by 0.2 dB at 20 dB, a difference smaller than the dispersion between realizations, and falls 3.0 dB below it at 25 dB} (Figure~\ref{fig:gain_methods}). \rev{The figure also includes the two other references, described in Section~\ref{sec:classical}: the Savitzky--Golay filter with a window of 5 samples and a polynomial of order 3, denoted SG(5,3), and the VisuShrink wavelet threshold computed with the exact noise level.} Inside the structural band \rev{the classical filters gain much less}: the low-pass and Savitzky--Golay filters \rev{gain at most 1.1 dB there}, VisuShrink with the exact noise level loses 2.2 dB \rev{at 15 dB input}, and TWN2N gains 3.3 dB at 20 dB and 5.7 to 8.2 dB from 15 dB down to 5 dB. Table~\ref{tab:bands} resolves the gains of every method of Figure~\ref{fig:gain_methods} by band: the full band, the structural band, and the two inter-resonance bands (V1 $=[1.15 f_1, 0.85 f_2]$, V2 $=[1.15 f_2, 0.85 f_3]$); the band SNR is computed from Tukey-windowed discrete Fourier transforms of the clean and processed position signals, summed over the three experiments and the three masses. At 15 dB input TWN2N gains 7.5 and 10.2 dB in the two inter-resonance bands, where the local SNR is \rev{low}, against 0.0 and 0.2 dB for the low-pass filter. The network therefore lowers the noise floor throughout the band instead of cutting it above $f_3$.

\begin{table}[h!]
\centering
\caption{SNR gains (positions, mean over 30 realizations) of every method of Figure~\ref{fig:gain_methods}, resolved by band: full band, structural band $0$--$1.25 f_3$, and inter-resonance bands V1 $=[1.15 f_1, 0.85 f_2]$ and V2 $=[1.15 f_2, 0.85 f_3]$. \rev{SG(5,3) is the Savitzky--Golay filter with a window of 5 samples and a polynomial of order 3.} The best method in each band and level is in bold.}
\label{tab:bands}
\resizebox{0.8\columnwidth}{!}{%
\begin{tabular}{cccccc}
\hline
\textbf{Input SNR} & \textbf{Method} & \textbf{Full band} & \textbf{$0$--$1.25 f_3$} & \textbf{V1} & \textbf{V2} \\
\textbf{(dB)} & & \textbf{(dB)} & \textbf{(dB)} & \textbf{(dB)} & \textbf{(dB)} \\ \hline
25 & TWN2N & -0.96 & -0.14 & \textbf{+4.14} & +3.83 \\
25 & \revblock{LP} & \revblock{+2.00} & \revblock{+0.68} & +0.00 & \revblock{+0.21} \\
25 & VisuShrink ($\hat\sigma$) & -2.79 & -4.77 & -4.78 & \textbf{+5.07} \\
25 & SG(5,3) & \revblock{\textbf{+2.25}} & \textbf{+0.77} & +0.04 & +0.82 \\
20 & TWN2N & \textbf{+3.11} & \textbf{+3.28} & \textbf{+5.97} & \textbf{+7.55} \\
20 & \revblock{LP} & \revblock{+2.92} & \revblock{+0.97} & +0.00 & \revblock{+0.21} \\
20 & VisuShrink ($\hat\sigma$) & -1.72 & -3.62 & -3.04 & +7.49 \\
20 & SG(5,3) & +2.83 & +0.98 & +0.04 & +0.82 \\
15 & TWN2N & \textbf{+6.01} & \textbf{+5.67} & \textbf{+7.48} & +10.22 \\
15 & \revblock{LP} & +3.25 & +1.06 & +0.00 & +0.21 \\
15 & VisuShrink ($\hat\sigma$) & -0.41 & -2.24 & -0.69 & \textbf{+10.73} \\
15 & SG(5,3) & +3.02 & +1.04 & +0.04 & +0.82 \\
12 & TWN2N & \textbf{+7.54} & \textbf{+6.80} & \textbf{+7.92} & +11.61 \\
12 & \revblock{LP} & +3.35 & +1.09 & +0.00 & +0.21 \\
12 & VisuShrink ($\hat\sigma$) & +0.50 & -1.29 & +1.06 & \textbf{+13.35} \\
12 & SG(5,3) & +3.09 & +1.07 & +0.04 & +0.82 \\
10 & TWN2N & \textbf{+8.10} & \textbf{+7.40} & \textbf{+8.29} & +11.81 \\
10 & \revblock{LP} & +3.39 & +1.09 & +0.00 & +0.21 \\
10 & VisuShrink ($\hat\sigma$) & +1.16 & -0.62 & +2.24 & \textbf{+15.33} \\
10 & SG(5,3) & +3.11 & +1.06 & +0.04 & +0.82 \\
5 & TWN2N & \textbf{+8.88} & \textbf{+8.16} & \textbf{+8.35} & +12.26 \\
5 & \revblock{LP} & +3.41 & +1.10 & +0.00 & +0.21 \\
5 & VisuShrink ($\hat\sigma$) & +2.88 & +1.15 & +5.30 & \textbf{+22.05} \\
5 & SG(5,3) & +3.12 & +1.07 & +0.04 & +0.82 \\
\hline
\end{tabular}%
}
\end{table}

\begin{figure}[h!]
    \centering
    \includegraphics[width=0.9\columnwidth]{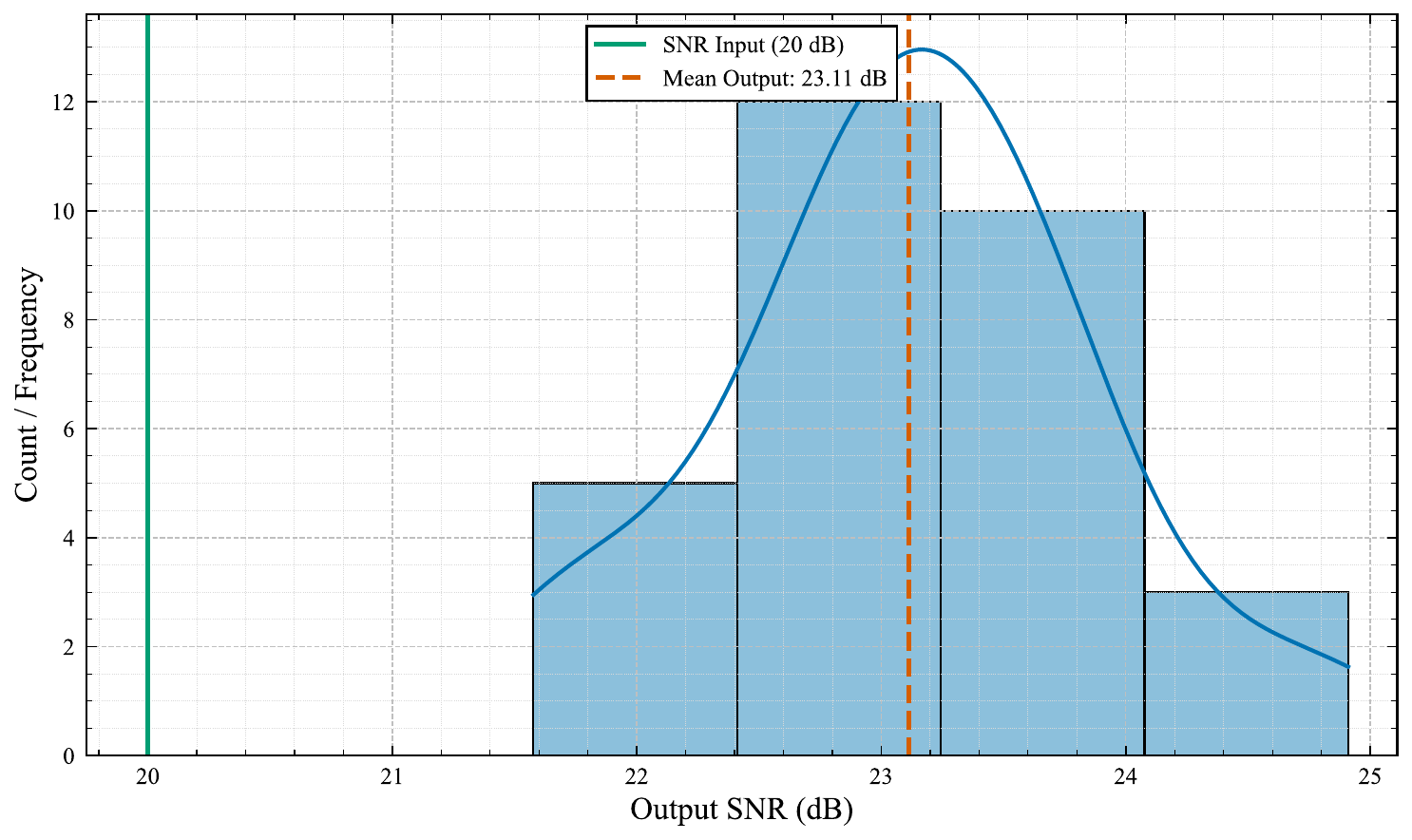}
    \caption{Histogram of the Output SNR for 30 independent runs with a target Input SNR of 20 dB.}
    \label{fig:histogram_20db}
\end{figure}

The distribution of the results can be further analyzed through histograms and boxplots. Figure \ref{fig:histogram_20db} presents the histogram of the Output SNR for the 20 dB input case across the 30 trials.

\begin{figure}[h!]
    \centering
    \includegraphics[width=0.9\columnwidth]{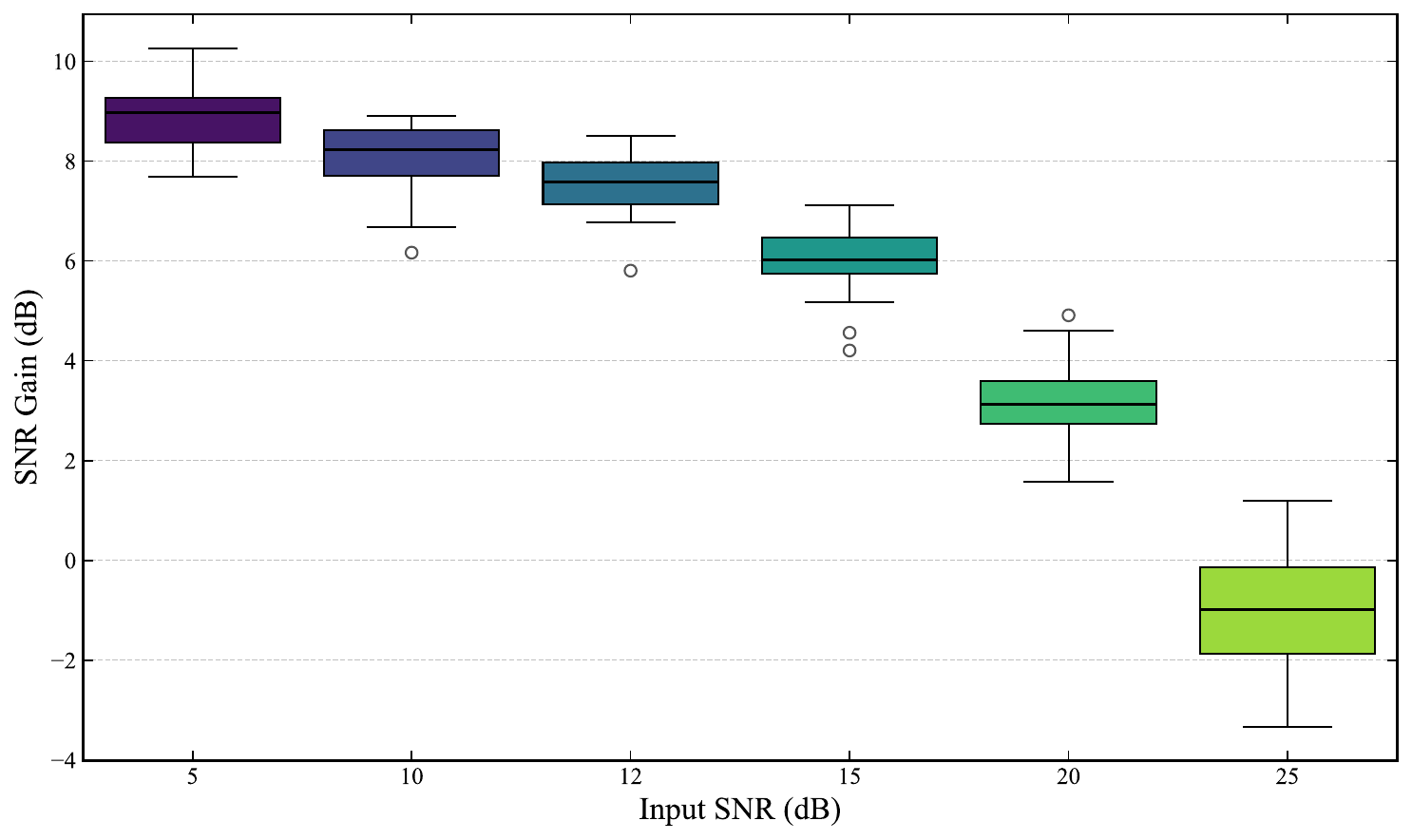}
    \caption{Boxplot of the SNR Gain distributions obtained from 30 experiments for each input noise level (5 dB to 25 dB).}
    \label{fig:boxplot_gain}
\end{figure}

Furthermore, Figure \ref{fig:boxplot_gain} illustrates the boxplot of the SNR Gains for all tested noise levels. The compact interquartile ranges evidence the reproducibility of the method at every level; the box at 25 dB lies entirely below zero, consistent with the upper bound of the operating range discussed above.

\subsubsection{Robustness to different noise distributions}
\label{sec:noise_types}

To further evaluate the robustness of the Time-Window Noise2Noise model, the model configured with a temporal window of $p=2$ and a latent space bottleneck of $\dim(r)=4$ was subjected to testing against various noise distributions. While the primary training and validation phases assumed additive Gaussian white noise, physical sensor data is frequently corrupted by colored noise or digitization artifacts. Therefore, the model's performance was assessed under three additional noise profiles: pink noise, brown noise, and quantization noise.

\rev{For stationary noise, the autocorrelation function and the power spectral density (PSD) form a Fourier-transform pair.} The characteristics of these noise distributions are defined as follows\rev{, each by its PSD and by the correlation between its samples}:

\begin{itemize}
    \item \rev{\textbf{White Noise:} The reference case, used in the other experiments of this work. Its PSD is flat up to the Nyquist frequency, and its samples are uncorrelated (zero correlation length).}
    \item \textbf{Pink Noise:} Characterized by a \rev{PSD} that is inversely proportional to the frequency, i.e., $S(f) \propto 1/f$. In the time domain, pink noise exhibits long-range temporal correlations. It is ubiquitous in physical systems, often representing fluctuations in electronic devices and sensors.
    \item \textbf{Brown Noise:} Also known as Brownian or red noise, its PSD is inversely proportional to the square of the frequency, $S(f) \propto 1/f^2$. It is generated by the integration of white noise, representing low-frequency dominant processes such as thermal drift or random walk dynamics. \rev{Its correlation is longer than that of pink noise and extends over the whole record.}
    \item \textbf{Quantization Noise:} Unlike the continuous stochastic processes above, quantization noise is a deterministic artifact arising from the Analog-to-Digital Conversion (ADC). It is mathematically modeled as the difference between the continuous input signal $X$ and its quantized discrete representation $X_q$:
    \begin{equation}
        \varepsilon_q = X_q - X = \Delta \cdot \text{round}\left(\frac{X}{\Delta}\right) - X~,
    \end{equation}
where $\Delta$ represents the quantization step size (resolution). \rev{Under high-resolution conditions, this error is often approximated as additive uniform white noise~\cite{oppenheim2010discrete}, an approximation that does not hold for the deliberately poorly ranged acquisition described below. Because it is a function of the signal, its PSD and its correlation are not fixed in advance: they depend on the signal and on the quantization step.}
\end{itemize}

For this robustness assessment, 30 independent trials were conducted for each noise profile. The continuous noises (white, pink, and brown) were calibrated to a target Input SNR of roughly 15 dB. The quantization noise experiment utilized a resolution $\Delta$ derived from an ADC configuration with a 12-bit resolution and a full-scale voltage range of 20 V (spanning from $-10$ V to $+10$ V); each channel was scaled so that its peak amplitude uses 0.35\% of the positive full scale (35 mV), i.e. a deliberately poorly ranged acquisition in which the rounding error is comparable to the signal, which yields an input SNR of approximately 15.4 dB.

\begin{table}[h!]
\centering
\caption{Statistical performance of the model ($p=2, \dim(r)=4$) across different noise distributions (30 independent trials; \rev{mean and standard deviation}), with the \rev{gain of the low-pass filter (cutoff at $1.05 f_3$)} and the in-band ($0$--$1.25 f_3$) gains.}
\label{tab:noise_types_robustness}
\resizebox{0.9\columnwidth}{!}{%
\begin{tabular}{lccccccc}
\hline
\textbf{Noise Type} & \textbf{Input SNR} & \textbf{Mean Output} & \textbf{Std Dev} & \textbf{Mean Gain} & \textbf{LP Gain} & \textbf{In-band Gain} & \textbf{In-band Gain} \\
& \textbf{(dB)} & \textbf{(dB)} & \textbf{(dB)} & \textbf{(dB)} & \textbf{(dB)} & \textbf{TWN2N (dB)} & \textbf{LP (dB)} \\ \hline
White & 15.00 & 21.01 & 0.68 & 6.01 & 3.25 & 5.67 & \revblock{1.06} \\
Pink & 15.00 & 17.48 & 0.66 & 2.48 & \revblock{0.39} & 2.86 & \revblock{0.14} \\
Brown & 15.00 & 15.28 & 0.50 & 0.28 & \revblock{-0.07} & 0.49 & \revblock{-0.02} \\
Quantization & 15.36 & 15.73 & 0.28 & 0.37 & 1.80 & 0.32 & \revblock{0.68} \\
\hline
\end{tabular}%
}
\end{table}

\begin{figure}[h!]
    \centering
    \includegraphics[width=\columnwidth]{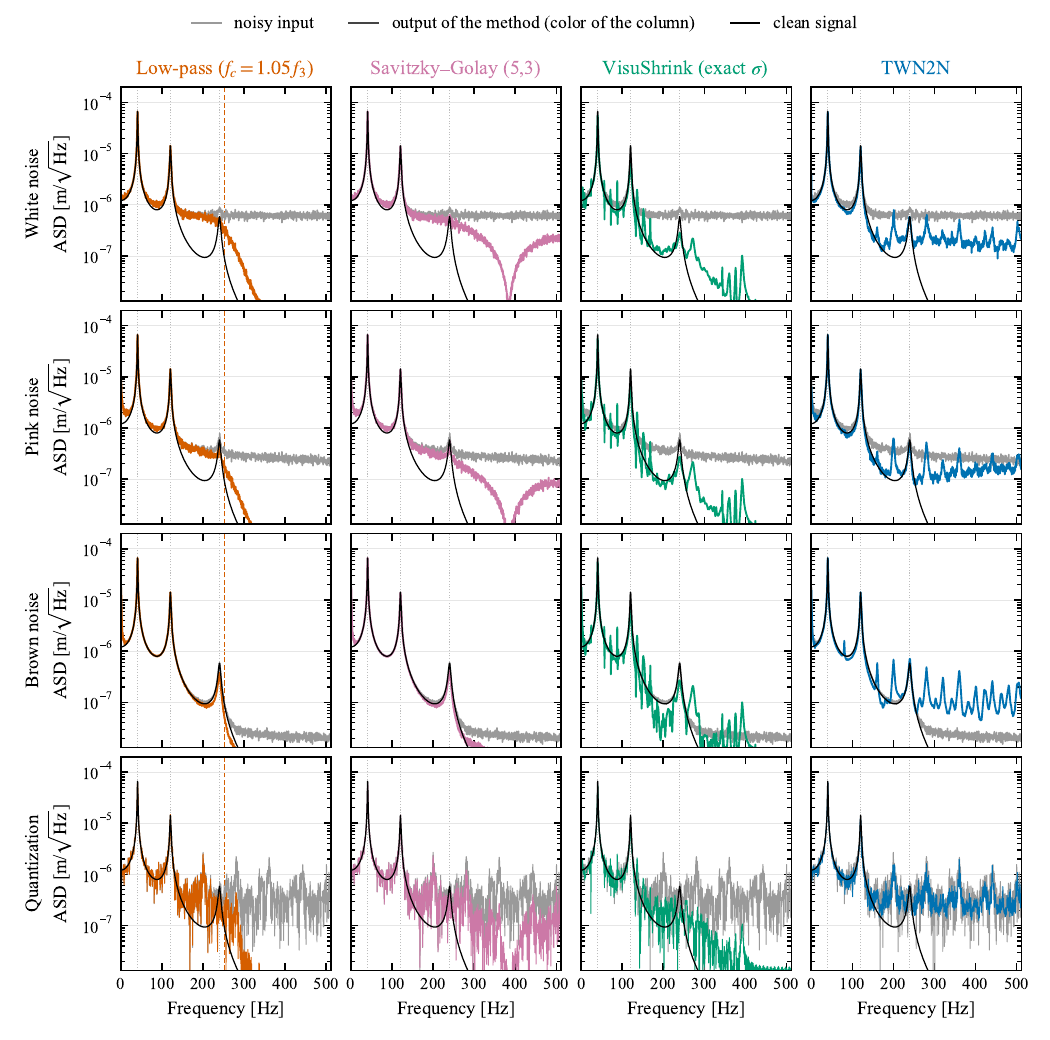}
    \caption{\rev{Amplitude spectral density (ASD) of the displacement $x_3$ for the impulse at $m_1$, at 15 dB input SNR (15.4 dB for quantization), for the four noise types (rows) and the four methods (columns): noisy input, output of the method, and clean signal. Each curve is the square root of the PSD averaged over the 30 realizations. The dotted lines mark the three natural frequencies, and the dashed line marks the cutoff of the low-pass filter}.}
    \label{fig:spectra}
\end{figure}

Table \ref{tab:noise_types_robustness} presents the \rev{aggregated results}. The results of Table~\ref{tab:noise_types_robustness} show that the gain depends on the spectral and temporal structure of the noise\rev{, and Figure~\ref{fig:spectra} shows the spectra of the displacement $x_3$ before and after each method. The network estimates each sample from its neighbors; for the power-law noises tested, the steeper the spectrum, the larger the part of the noise that the neighbors predict. Under white noise, the low-pass and Savitzky--Golay filters remove the noise above the third resonance and leave the floor between the resonances unchanged, VisuShrink lowers the second and third resonance peaks to about half of their clean height, and TWN2N lowers the floor over the whole band with the three peaks within 5\% of their clean height.} Under pink noise the gain falls from 6.0 dB (white) to 2.5 dB, whereas the low-pass filter gains only 0.4 dB, because most of the $1/f$ noise power lies inside the structural band; there, TWN2N gains 2.9 dB against \rev{0.1 dB} for the filter.

Under brown noise the gain is $0.3 \pm 0.5$ dB\rev{, negligible in size}. The $1/f^2$ noise is strongly correlated in time, so the noise value at the masked instant is largely predictable from its neighbors and the blind-spot construction no longer separates it from the deterministic dynamics: the conditional \rev{median} of the target given the window \rev{then includes the predictable part of the noise, and the independence condition stated in Section~\ref{sec:architecture} no longer holds}. This is a limitation of the method for temporally correlated noise. \rev{For the linear filters, the limit is the overlap with the structural band:} 99\% of the brown noise power lies below the first natural frequency\rev{, inside the band that the low-pass filter preserves}. None of the classical filters gains anything either (\rev{Tables~\ref{tab:noise_types_robustness} and~\ref{tab:sg_colored}}): the low-pass filter \revblock{leaves the SNR practically unchanged ($-0.1$ dB)} and the wavelet filter degrades the signal.

Under quantization noise, with the deliberately poorly ranged 12-bit acquisition described above (input SNR 15.4 dB), the gain is 0.4 dB, lower than that of the low-pass filter (1.8 dB) and of the Savitzky--Golay filter (1.8 dB). The rounding error is \rev{deterministic, bounded, and signal-dependent, with a broadband but not white spectrum in this acquisition}, and at this level it is close to the reconstruction floor of the network; the architecture offers no advantage for this noise type.

\rev{The spectra of Figure~\ref{fig:spectra} also show peaks in the output of TWN2N that the clean signal does not contain, at frequencies near integer combinations of the natural frequencies (200, 280, and 360 Hz, for example), consistent with harmonic and intermodulation distortion introduced by the nonlinear mapping. They lie about 40 dB below the first resonance, at the level of the noise floor of the white-noise input, and they stand above the input where the noise is weaker than these peaks, as under brown noise above about 250 Hz. They are not structural modes.}

Taken together, the method is most effective for noise that is uncorrelated in time (white)\rev{, keeps part of its gain under pink noise,} gives no meaningful gain for strongly correlated noise \rev{(brown)}, and is not the method of choice for quantization noise at this level. \rev{Under pink noise, it is also the method with the largest gain among those compared: 2.5 dB against 1.3 dB for the best wavelet setting and 0.4 dB for the low-pass and Savitzky--Golay filters (Tables~\ref{tab:noise_types_robustness} and~\ref{tab:sg_colored}).} The band-resolved gains of Table~\ref{tab:bands} \rev{(white noise) and the spectra of Figure~\ref{fig:spectra}} show the same behavior in the frequency domain: the network lowers the noise floor \rev{throughout the band when} the noise is uncorrelated, \rev{lowers it less as the correlation grows, and does not lower it when the correlation is strong}. The scope of the method is therefore bounded by the independence assumption\rev{: for signal-independent noise, performance deteriorates as the noise becomes more predictable from its temporal neighborhood, a property set by its correlation and spectral structure, whereas the overlap of the noise spectrum with the structural band limits the linear filters}.

\subsubsection{Identification of natural frequencies and damping ratios}
\label{sec:modal_id}
The natural frequencies and damping ratios of the 3-DOF system were identified from the noisy and from the denoised signals to assess the practical impact of denoising on system characterization. The identification was conducted on both raw noisy signals and signals reconstructed by the denoising model at two distinct noise levels (20 dB and 15 dB). For this purpose, a classical time-domain technique was employed: the Ibrahim Time Domain (ITD) method \cite{ibrahim1977method}. The theoretical background for this estimator is detailed in Ewins \cite{ewins2009modal}.

First, the theoretical natural frequencies and damping ratios of the simulated system were calculated to serve as the ground truth for validation. These target values are presented in Table \ref{tab:theoretical_params}.

\begin{table}[h!]
\centering
\caption{Theoretical (target) natural frequencies and damping ratios of the 3-DOF system.}
\label{tab:theoretical_params}
\resizebox{0.5\columnwidth}{!}{%
\begin{tabular}{cccc}
\hline
\textbf{Mode} & \textbf{$\omega_n$ (rad/s)} & \textbf{$f_n$ (Hz)} & \textbf{$\zeta$} \\ \hline
1 & 252.26 & 40.15 & 0.0100 \\
2 & 754.62 & 120.10 & 0.0072 \\
3 & 1507.40 & 239.91 & 0.0100 \\
\hline
\end{tabular}%
}
\end{table}

Table~\ref{tab:itd_comparison} presents the estimates obtained with the ITD method on one realization, and Table~\ref{tab:uncertainty_budget} the ensemble statistics over the thirty realizations at 20 and 15 dB, \rev{before and after denoising; the low-pass values are quoted below}. The ITD is applied to the displacements of the three masses for the impulse at $m_1$ with a block delay of one sample and an oversized model order of 20, the setting that recovers the three modes exactly from the clean signal and most robustly from the noisy one; each identified mode is the pole nearest to the reference frequency within $\pm 5\%$, and poles with $\zeta \ge 0.5$, which the oversized model produces to absorb noise, are discarded as spurious. Because the system is lightly damped and the natural frequencies are well separated, the frequencies of the first two modes are recovered from the raw noisy signals within 0.1\% and their damping ratios within 2--19\%; denoising changes little for these modes, and the run-to-run dispersion of their damping ratios is in fact larger after denoising ($s = 0.0005$ against 0.0001 for the first mode at 20 dB), because the residual of the network is not white and perturbs the decay envelope slightly. One realization at 15 dB gives $\zeta_2 = 0.062$ from the denoised signal, against 0.0058 to 0.0072 (median 0.0062) in the other 29: the pole sits at the right frequency but the estimator misreads its decay on that realization; the value is kept in Table~\ref{tab:uncertainty_budget}, where it dominates the dispersion of that entry, and the median is quoted here for that reason. The third mode, whose contribution to the displacements is about 35 times smaller than that of the first mode and which decays within 0.1 s, is where the noise matters: from the raw signals the ITD estimates its damping ratio as $0.054 \pm 0.005$ (20 dB) and $0.084 \pm 0.012$ (15 dB, mean $\pm$ standard deviation) against a reference of 0.010, with a standard deviation of 9--25 rad/s on its frequency. From the denoised signals the same estimator gives $0.0093 \pm 0.0007$ and $0.0099 \pm 0.0011$, with a frequency dispersion of about 1 rad/s. The 95\% confidence intervals of the mean, [0.0091, 0.0096] at 20 dB and [0.0095, 0.0103] at 15 dB, show that the denoised estimate is biased low by 7\% at 20 dB and contain the reference at 15 dB: the denoising removes the several-fold overestimation but leaves a residual bias of the order of the reconstruction floor. The low-pass filter, whose cutoff \rev{lies} just above the third resonance, biases that mode \revblock{($1524 \pm 4$ rad/s and $\zeta = 0.022 \pm 0.002$ at 20 dB, $1531 \pm 4$ rad/s and $\zeta = 0.030 \pm 0.004$ at 15 dB, not tabulated)}: a linear filter cannot lower the noise floor around a resonance without distorting it, whereas the network reduces the floor throughout the band.

\begin{table}[h!]
\centering
\caption{Natural frequencies and damping ratios identified with the ITD from the noisy and from the denoised signals (20 and 15 dB input SNR, one realization).}
\label{tab:itd_comparison}
\resizebox{0.80\columnwidth}{!}{%
\begin{tabular}{ccccccc}
\hline
\multirow{2}{*}{\textbf{Input SNR}} & \multirow{2}{*}{\textbf{Mode}} & \multicolumn{2}{c}{\textbf{Noisy Input}} & & \multicolumn{2}{c}{\textbf{Denoised Output}} \\ \cline{3-4} \cline{6-7}
 &  & \textbf{$\omega_n$ (rad/s)} & \textbf{$\zeta$} & & \textbf{$\omega_n$ (rad/s)} & \textbf{$\zeta$} \\ \hline
\multirow{3}{*}{20 dB} & 1 & 252.28 & 0.010 & & 252.28 & 0.010 \\
 & 2 & 754.67 & 0.008 & & 754.64 & 0.007 \\
 & 3 & 1506.55 & 0.065 & & 1504.20 & 0.010 \\
\hline
\multirow{3}{*}{15 dB} & 1 & 252.26 & 0.011 & & 252.21 & 0.010 \\
 & 2 & 754.57 & 0.008 & & 754.68 & 0.006 \\
 & 3 & 1499.14 & 0.074 & & 1507.34 & 0.011 \\
\hline
\end{tabular}%
}
\end{table}

The ensemble statistics of Table~\ref{tab:uncertainty_budget} use the $N = 30$ independent Monte Carlo realizations of Section~\ref{Uncertainty}: each natural frequency and damping ratio is characterized by its ensemble mean and sample standard deviation $s$, the run-to-run dispersion of a single identification, and the table also gives the number of realizations in which each mode was identified. Statements about the mean use the 95\% confidence interval of Section~\ref{Uncertainty}, whose half-width is $t_{0.975,29}\,s/\sqrt{N} = 0.37\,s$. The dispersions are small enough for the differences discussed above to be resolved: the mean damping ratio of the third mode from the raw signals differs from the reference by 8.5 (20 dB) and 6.1 (15 dB) standard deviations, i.e. by 23 and 16 confidence-interval half-widths.

\begin{table}[h!]
\centering
\caption{Ensemble statistics of the ITD-identified natural frequencies (rad/s) and damping ratios over $N=30$ Monte Carlo realizations, before (noisy) and after (TWN2N) denoising, at 20 and 15 dB input SNR (the corresponding values after the low-pass filter are quoted in the text). Values are the ensemble mean $\pm$ sample standard deviation $s$ (run-to-run dispersion; the 95\% confidence interval of the mean has half-width $0.37\,s$); the count in parentheses is the number of realizations in which the mode was identified.}
\label{tab:uncertainty_budget}
\resizebox{\columnwidth}{!}{%
\begin{tabular}{cccccccc}
\hline
\textbf{Input} & \textbf{Mode} & \textbf{Ref.} & \textbf{Ref.} & \textbf{Noisy} & \textbf{Noisy} & \textbf{TWN2N} & \textbf{TWN2N} \\
\textbf{SNR} & & $\omega_n$ & $\zeta$ & $\omega_n \pm s$ (found) & $\zeta \pm s$ & $\omega_n \pm s$ (found) & $\zeta \pm s$ \\ \hline
\multirow{3}{*}{20 dB} & 1 & 252.3 & 0.0100 & 252.3 $\pm$ 0.0 (30/30) & 0.0102 $\pm$ 0.0001 & 252.2 $\pm$ 0.1 (30/30) & 0.0094 $\pm$ 0.0005 \\
 & 2 & 754.6 & 0.0072 & 754.6 $\pm$ 0.1 (29/30) & 0.0076 $\pm$ 0.0001 & 754.6 $\pm$ 0.1 (30/30) & 0.0065 $\pm$ 0.0004 \\
 & 3 & 1507.4 & 0.0100 & 1507.7 $\pm$ 8.9 (30/30) & 0.0545 $\pm$ 0.0053 & 1506.7 $\pm$ 1.0 (30/30) & 0.0093 $\pm$ 0.0007 \\
\hline
\multirow{3}{*}{15 dB} & 1 & 252.3 & 0.0100 & 252.2 $\pm$ 0.0 (30/30) & 0.0106 $\pm$ 0.0001 & 252.2 $\pm$ 0.1 (30/30) & 0.0095 $\pm$ 0.0004 \\
 & 2 & 754.6 & 0.0072 & 754.6 $\pm$ 0.2 (30/30) & 0.0085 $\pm$ 0.0002 & 754.7 $\pm$ 0.2 (30/30) & 0.0081 $\pm$ 0.0102 \\
 & 3 & 1507.4 & 0.0100 & 1508.2 $\pm$ 25.0 (30/30) & 0.0844 $\pm$ 0.0123 & 1507.1 $\pm$ 1.0 (30/30) & 0.0099 $\pm$ 0.0011 \\
\hline
\end{tabular}%
}
\end{table}

\subsubsection{Sensitivity to the damping level}
\label{sec:damping_ablation}
Because the damping ratio governs both the length of the free decay and the difficulty of estimating $\zeta$ from a noisy record, the study was repeated at 15 dB input SNR for seven proportional-damping configurations of the same $\mathbf{M}$ and $\mathbf{K}$, Equation~(\ref{eq:rayleigh}), listed in the first two columns of Table~\ref{tab:damping}: Rayleigh damping with $\zeta_1 = \zeta_3$ equal to 0.002 (i), 0.005 (ii), 0.010 (iii, the main configuration of the paper), 0.020 (iv), and 0.050 (v); mass-proportional damping only (vi: $\beta = 0$ and $\zeta_1 = 0.02$, hence $\zeta = 0.020/0.007/0.003$); and stiffness-proportional damping only (vii: $\alpha = 0$ and $\zeta_3 = 0.02$, hence $\zeta = 0.003/0.010/0.020$). The record length follows the decay of the slowest mode in each case (seven time constants, clipped to 1--8 s): 8 s for (i), 6 s for (ii), 3 s for (iii), 1.5 s for (iv) and (vi), 1 s for (v), and 8 s for (vii).

\begin{figure}[h!]
    \centering
    \includegraphics[width=0.98\columnwidth]{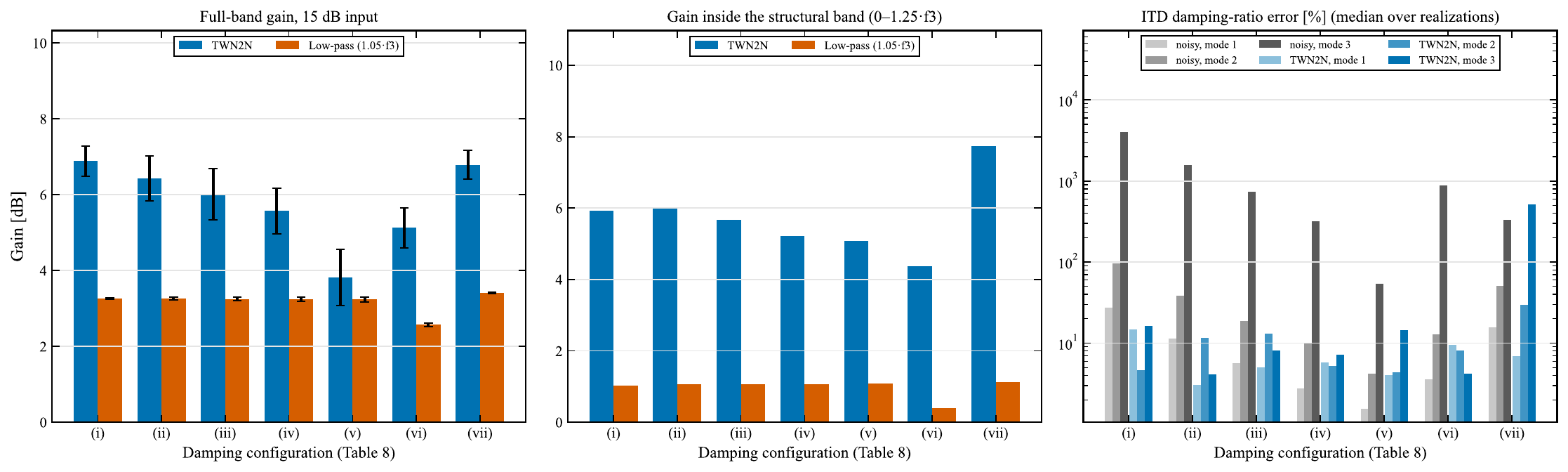}
    \caption{Damping sensitivity at 15 dB input SNR: full-band gain of TWN2N and of the low-pass filter \rev{with cutoff at $1.05 f_3$} (left), gain inside the structural band (center) and ITD damping-ratio error of the identified modes before and after denoising (right).}
    \label{fig:damping}
\end{figure}

Table~\ref{tab:damping} and Figure~\ref{fig:damping} report, for thirty realizations per configuration, the full-band and in-band SNR gains of TWN2N and of the low-pass filter and the ITD identification errors before and after denoising (median over the realizations in which the mode was identified). All three modes were identified in every realization, except mode 3 of (v), missed in one noisy realization and one denoised realization; mode 1 of (vi), missed in one noisy realization; mode 3 of (vii), missed in 18 denoised realizations. The gain \rev{persists across the damping levels}: for Rayleigh damping between 0.002 and 0.020, configurations (i) to (iv), TWN2N gains 5.6--6.9 dB in the full band and 5.2--6.0 dB inside the structural band, against \rev{3.3 dB and 1.0--1.1 dB} for the low-pass filter, the largest \rev{full-band gain} occurring for the lightest damping, where the record contains more cycles of every mode. At $\zeta = 0.05$ (v) the record lasts only 1 s and the gain falls to 3.8 dB, still above the low-pass filter. Mass-proportional (vi) and stiffness-proportional (vii) damping give 5.1 and 6.8 dB. The identification errors follow the pattern of Section~\ref{sec:modal_id}: the frequencies are recovered from the raw signals in every configuration, but the damping ratio of the third mode is overestimated from the noisy signals by 54\% to 40-fold, and \rev{differs from its reference by only} 4--16\% after denoising, for configurations (i) to (vi). The exception is the stiffness-proportional configuration (vii), in which the third mode ($\zeta_3 = 0.02$) decays within 0.05 s of an 8 s record dominated by the ringing of the first mode ($\zeta_1 = 0.003$): its short transient is not reconstructed by the network, the mode is identified in 12 of the 30 denoised realizations with a median damping error of 519\%, and denoising does not help its estimation. The record must therefore contain enough samples of the decay of each mode of interest, with that mode above the noise floor; when the modal decay rates differ by more than an order of magnitude, a single record length cannot satisfy this condition for every mode, and the identification of the fast mode is lost.

\begin{table}[h!]
\centering
\caption{Sensitivity to the damping level at 15 dB input SNR (30 realizations per configuration). The configurations (i) to (vii) are defined in the text; (iii)$^{*}$ is the main configuration of the paper. ITD columns give the median errors (modes 1/2/3) of the natural frequencies and damping ratios identified from the noisy and from the denoised signals; the number of realizations in which each mode was identified is given in the text.}
\label{tab:damping}
\resizebox{\columnwidth}{!}{%
\begin{tabular}{lcccccccc}
\hline
\multirow{2}{*}{\textbf{Config.}} & \multirow{2}{*}{\textbf{$\zeta_1/\zeta_2/\zeta_3$}} & \textbf{Gain TWN2N} & \textbf{Gain LP} & \textbf{In-band gain} & \multicolumn{2}{c}{\textbf{ITD noisy}} & \multicolumn{2}{c}{\textbf{ITD TWN2N}} \\ \cline{6-7} \cline{8-9}
 & & \textbf{(dB)} & \textbf{(dB)} & \textbf{TWN2N / LP (dB)} & \textbf{err $\omega_n$ (\%)} & \textbf{err $\zeta$ (\%)} & \textbf{err $\omega_n$ (\%)} & \textbf{err $\zeta$ (\%)} \\ \hline
(i) & 0.0020 / 0.0014 / 0.0020 & 6.89 $\pm$ 0.40 & 3.26 & 5.93 / \revblock{1.03} & 0.01/0.01/0.73 & 28/96/4047 & 0.00/0.00/0.01 & 15/5/16 \\
(ii) & 0.0050 / 0.0036 / 0.0050 & 6.43 $\pm$ 0.60 & 3.26 & 6.02 / \revblock{1.07} & 0.01/0.01/0.77 & 12/39/1568 & 0.01/0.01/0.02 & 3/12/4 \\
(iii)$^{*}$ & 0.0100 / 0.0072 / 0.0100 & 6.01 $\pm$ 0.67 & 3.25 & 5.67 / \revblock{1.06} & 0.01/0.01/1.30 & 6/19/731 & 0.03/0.01/0.04 & 5/13/8 \\
(iv) & 0.0200 / 0.0143 / 0.0200 & 5.58 $\pm$ 0.60 & 3.25 & 5.22 / \revblock{1.06} & 0.03/0.03/1.15 & 3/10/317 & 0.04/0.04/0.10 & 6/5/7 \\
(v) & 0.0500 / 0.0358 / 0.0500 & 3.82 $\pm$ 0.75 & 3.24 & 5.07 / \revblock{1.09} & 0.05/0.08/1.53 & 2/4/54 & 0.14/0.15/0.56 & 4/4/15 \\
(vi) & 0.0200 / 0.0067 / 0.0033 & 5.13 $\pm$ 0.53 & \revblock{2.57} & 4.38 / \revblock{0.41} & 0.02/0.02/0.23 & 4/13/887 & 0.05/0.01/0.02 & 10/8/4 \\
(vii) & 0.0033 / 0.0100 / 0.0200 & 6.79 $\pm$ 0.38 & 3.41 & 7.74 / \revblock{1.12} & 0.01/0.03/1.99 & 16/51/330 & 0.02/0.03/2.55 & 7/30/519 \\
\hline
\end{tabular}%
}
\end{table}

\subsubsection{Comparison with classical filters}
\label{sec:classical}

To contextualize the performance of the proposed Time-Window Noise2Noise (TWN2N) model, a comparative analysis was conducted against the Savitzky--Golay filter and against the VisuShrink wavelet threshold, the latter tuned with the exact noise level, which is unavailable in practice. The comparison with the low-pass filter was presented in Section~\ref{sec:3dof_results} (Tables~\ref{tab:snr_lowpass} and~\ref{tab:bands}) and is not repeated here. These references serve to evaluate the trade-offs between the proposed approach and deterministic signal processing methods.

\paragraph{Savitzky--Golay filter}

The Savitzky--Golay (SG) filter \cite{savitzky1964smoothing} is a widely adopted digital filter that smooths data by fitting successive sub-sets of adjacent data points with a polynomial from least squares optimization. For a direct comparison with the TWN2N model, the SG filter was configured with a polynomial order of 3 and two window sizes: 5 samples, which yields a temporal receptive field comparable to that of the proposed method ($p=2$, comprising 4 neighboring points), and 15 samples, a wider window of the kind commonly adopted in practice\rev{; the two configurations are denoted SG(5,3) and SG(15,3)}. A crucial distinction must be highlighted: the SG filter utilizes the instantaneous central frame in its estimation process, whereas the TWN2N architecture is strictly blind to the central frame to prevent trivial identity mapping.

Tables~\ref{tab:sg_white} and~\ref{tab:sg_colored} gather the output SNR of the Savitzky--Golay and wavelet filters and of TWN2N, for white noise and for colored and quantization noise, respectively, on the same noisy realizations.

Table~\ref{tab:sg_white} shows that the SG filter with window 5 gains 2.3--3.1 dB at every level, essentially the removal of the noise above the structural band by a short smoother (1.0 dB gain inside the structural band, Table~\ref{tab:bands}); TWN2N exceeds it from 20 dB down (3.1 against 2.8 dB at 20 dB, 6.0 against 3.0 dB at 15 dB and 8.9 against 3.1 dB at 5 dB) and falls below it at 25 dB. The wider window (15 samples, 0.015 s) spans more than three periods of the third mode, \rev{removes the second mode and most of the third, and limits the output SNR to 9.9--12.6 dB regardless of the input, a loss from 12 dB input up}. Under colored noise (Table~\ref{tab:sg_colored}) the SG filter is neutral for pink ($+0.4$ dB) and brown ($-0.05$ dB) noise and gains 1.8 dB for quantization noise, where it outperforms TWN2N.

\begin{table}[h!]
\centering
\caption{Output SNR of the classical filters and of TWN2N under white noise (mean over the same 30 noisy realizations): Savitzky--Golay (windows 5 and 15, order 3) and Symlet-8 VisuShrink with $\hat\sigma = \sigma_{\text{true}}/3$, $\sigma_{\text{true}}$, and $3\sigma_{\text{true}}$.}
\label{tab:sg_white}
\resizebox{0.8\columnwidth}{!}{%
\begin{tabular}{ccccccc}
\hline
\textbf{Input SNR} & \textbf{SG (5,3)} & \textbf{SG (15,3)} & \textbf{Visu $\hat\sigma/3$} & \textbf{Visu $\hat\sigma$} & \textbf{Visu $3\hat\sigma$} & \textbf{TWN2N} \\
\textbf{(dB)} & \textbf{(dB)} & \textbf{(dB)} & \textbf{(dB)} & \textbf{(dB)} & \textbf{(dB)} & \textbf{(dB)} \\ \hline
25 & 27.25 & 12.55 & 28.23 & 22.21 & 15.00 & 24.04 \\
20 & 22.83 & 12.47 & 24.02 & 18.28 & 11.53 & 23.11 \\
15 & 18.02 & 12.23 & 19.89 & 14.59 & 8.26 & 21.01 \\
12 & 15.09 & 11.90 & 17.44 & 12.50 & 6.35 & 19.54 \\
10 & 13.11 & 11.54 & 15.80 & 11.16 & 5.13 & 18.09 \\
5 & 8.12 & 9.87 & 11.81 & 7.88 & 2.52 & 13.87 \\
\hline
\end{tabular}%
}
\end{table}

\begin{table}[h!]
\centering
\caption{Output SNR of the classical filters and of TWN2N under colored and quantization noise (mean over the same 30 realizations); columns as in Table~\ref{tab:sg_white}.}
\label{tab:sg_colored}
\resizebox{0.85\columnwidth}{!}{%
\begin{tabular}{lccccccc}
\hline
\textbf{Noise Type} & \textbf{Input SNR} & \textbf{SG (5,3)} & \textbf{SG (15,3)} & \textbf{Visu $\hat\sigma/3$} & \textbf{Visu $\hat\sigma$} & \textbf{Visu $3\hat\sigma$} & \textbf{TWN2N} \\
& \textbf{(dB)} & \textbf{(dB)} & \textbf{(dB)} & \textbf{(dB)} & \textbf{(dB)} & \textbf{(dB)} & \textbf{(dB)} \\ \hline
White & 15.00 & 18.02 & 12.23 & 19.89 & 14.59 & 8.26 & 21.01 \\
Pink & 15.00 & 15.41 & 11.04 & 16.28 & 13.26 & 7.93 & 17.48 \\
Brown & 15.00 & 14.95 & 10.62 & 14.32 & 11.90 & 7.46 & 15.28 \\
Quantization & 15.36 & 17.12 & 11.92 & 17.39 & 14.57 & 8.62 & 15.73 \\
\hline
\end{tabular}%
}
\end{table}

\paragraph{Symlet VisuShrink filter}

Wavelet-based denoising represents another cornerstone of state-of-the-art signal filtering. In conventional thresholding schemes, a global (universal) threshold is commonly used to filter small wavelet coefficients. However, this procedure can inadvertently remove high-frequency components, such as sharp transition edges. To improve upon simple coefficient truncation, soft-thresholding \cite{donoho1995noising} reconstructs the signal by translating the empirical wavelet coefficients towards zero by a specific threshold amount, rather than simply discarding them. This approach effectively dampens the small coefficients considered to be noise while preserving the strength of active structural elements. Advanced multiscale implementations often expand on this by partitioning data to exploit local noise characteristics and compute subband-dependent thresholds for precise edge localization without undesirable boundary effects. For this evaluation, the highly optimized Symlet wavelet family \cite{daubechies1992ten} was employed in conjunction with the VisuShrink technique \cite{donoho1994ideal} using the foundational ``soft'' thresholding mode.

The universal threshold $\lambda$ in VisuShrink is mathematically defined as:
\begin{equation}
    \lambda = \hat{\sigma} \sqrt{2 \ln(n)}~,
\end{equation}
where $\hat{\sigma}$ is the estimated standard deviation of the noise and $n$ is the number of samples in the signal. A significant practical limitation of this method is its heavy reliance on $\hat{\sigma}$, an intrinsic noise property that is strictly unknown in real-world scenarios. Defining $\sigma_{\text{true}}$ as the real standard deviation of the noise, practical applications inevitably require its estimation ($\hat{\sigma} \approx \sigma_{\text{true}}$). To simulate this uncertainty, the TWN2N model was compared against the Symlet VisuShrink filter under three threshold scenarios: the exact value ($\hat{\sigma} = \sigma_{\text{true}}$), which is unattainable in practice; a threefold overestimation ($\hat{\sigma}=3\sigma_{\text{true}}$), emulating a practical condition where the noise statistics are poorly characterized, and the more consequential form of misestimation, since it leads to aggressive over-filtering that suppresses valid signal components alongside the noise; and a threefold underestimation ($\hat{\sigma}=\sigma_{\text{true}}/3$), a milder threshold.

Tables~\ref{tab:sg_white} and~\ref{tab:sg_colored} summarize the results for white and colored noises, respectively. For this system the universal threshold with the exact noise level degrades the signal \rev{from 15 dB input up} (output SNR of 14.6 dB at 15 dB input; $-2.2$ dB inside the structural band, Table~\ref{tab:bands})\rev{ and remains below TWN2N at every level}. VisuShrink is derived for signals that are sparse in the wavelet basis under zero-mean white Gaussian noise of known variance; with $f_3 = 0.47\,f_s/2$, the impulse responses are not sparse, the detail coefficients of the finest scales carry the third mode and part of the second, and the threshold $\lambda = \hat{\sigma}\sqrt{2\ln n}$ removes them together with the noise. A threefold overestimation of $\hat\sigma$ destroys the signal (8.3 dB), whereas a threshold three times smaller than the universal one is the best wavelet setting: it is above TWN2N at 25 and 20 dB input (28.2 and 24.0 dB against 24.0 and 23.1 dB) and below it from 15 dB down (19.9 against 21.0 dB at 15 dB, 15.8 against 18.1 dB at 10 dB and 11.8 against 13.9 dB at 5 dB). The sensitivity to $\hat{\sigma}$ and to the sparsity assumption reflects the limitation of threshold-based methods: their performance is conditioned on prior knowledge that is unavailable in real experimental scenarios, while TWN2N operates blindly.

\rev{For the other noise types (Table~\ref{tab:sg_colored}), with} the exact noise level the wavelet filter loses 1.7 dB under pink noise, 3.1 dB under brown noise and 0.8 dB under quantization noise; with the reduced threshold it gains 1.3 dB under pink noise, less than the 2.5 dB of TWN2N, loses 0.7 dB under brown noise, where no method \rev{gives a meaningful gain}, and gains 2.0 dB under quantization noise, where it outperforms TWN2N together with the Savitzky--Golay and low-pass filters.

\subsection{Noise suppression on the nonlinear impact dataset}
\label{sec:impact}
For the impact experiment dataset, the architecture was tested on the nonlinear force profile. The input vector $X(t_i)$ comprised $v_{\text{approach}}$, $e$, $\delta(t_i)$, $\dot{\delta}(t_i)$, and $F_N(t_i)$, with window size $p=2$ (Equation~(\ref{Eq_p})). The latent space dimension was set to 3, and training was regulated by the early-stopping rule of Appendix~A. The dataset comprises twenty impacts (five approach velocities between 0.1 and 0.5 m/s combined with four restitution coefficients between 0.76 and 0.97), simulated with the parameters of Section~\ref{sec:impact_dataset} and concatenated into one record of 299,522 samples on which a single network is trained.

The nominal restitution parameter $e$ of the contact law and the approach velocity $v_{\text{approach}}$ enter the network as known conditioning metadata: they are constant within each impact and are given without noise. In a real test $v_{\text{approach}}$ is measured and $e$ is not directly available, so the results of this section hold for a method that knows the test conditions; an ablation without the two metadata inputs is reported at the end of the section. The performance in the impact dataset was evaluated by signal error, by physical consistency, and by statistical significance across different noise regimes. Noise was added as a percentage of the signal's standard deviation (3\% to 15\%), resulting in SNRs of 30.46, 26.02, 23.10, 20.00, and 16.48 dB. To quantify the proximity of the signals to the ground truth, a state vector $\mathbf{v}(t_i) = [\delta(t_i), \dot{\delta}(t_i), F_N(t_i)]$ was defined and a dimensionless distance $d_i$ between the estimated and the clean state was computed, each component being normalized by the standard deviation $s_j$ of the corresponding clean signal over the dataset, as defined in Equation~(\ref{eq:euclidean_dist}):
\begin{equation}
    d_i = \sqrt{\frac{1}{3}\left[\left(\frac{\delta_{c} - \delta_{est}}{s_\delta}\right)^2 + \left(\frac{\dot{\delta}_{c} - \dot{\delta}_{est}}{s_{\dot\delta}}\right)^2 + \left(\frac{F_{N,c} - F_{N,est}}{s_F}\right)^2\right]}~,
    \label{eq:euclidean_dist}
\end{equation}
where the subscript $est$ refers to either the raw noisy measurement or the denoised model output; without the normalization the three terms have different units and the velocity and force errors dominate the sum.

The unit of replication is the trial: one noise realization and one training of the network on the record of the twenty impacts, repeated with three independent trials per noise level. Within a trial the twenty impacts are blocks (fixed combinations of five approach velocities and four restitution coefficients, processed by the same network), not independent replicates. The distance $d_i$ and the normalized root-mean-square error of each state variable (NRMSE, the root-mean-square error, RMSE, divided by the standard deviation $s_j$ of the clean signal) are therefore \rev{computed as root mean squares over the samples of each impact and then averaged} over the twenty impacts of the trial, and the effect of denoising is tested with a paired one-tailed $t$-test over the three trial means (2 degrees of freedom), the null hypothesis being that the mean distance of the denoised output is not smaller than that of the noisy input. The number of impacts in which the output is closer to the ground truth than the input in every trial is reported as a blocked summary. The samples are those of the clean contact interval ($\delta > 0$ in the theoretical solution), the same for both signals; the fraction of samples in which the noisy or the denoised signal violates the physical constraints $\delta \ge 0$ and $F_N \ge 0$ is reported separately instead of being excluded.

Table~\ref{tab:impact_distance} presents the results for the full state vector. At every noise level the denoised output is closer to the ground truth than the noisy input in all 20 impacts in every trial, the distance falls by 44\% to 66\% across levels and trials, and the trial-level test rejects the null hypothesis at every level ($p$-value $< 10^{-3}$ with 2 degrees of freedom); the within-trial paired tests over the twenty impacts give $t$ between -8 and -6, reported only as a blocked summary. Without the metadata inputs $e$ and $v_{\text{approach}}$ (last column of Table~\ref{tab:impact_distance}, 3 trials per level), the denoised distance is 0.018 to 0.056 against 0.013 to 0.042 with them, i.e. the network still reduces the distance by 25\% to 53\%; the conditioning inputs account for the remainder.

\begin{table}[h!]
\centering
\caption{Dimensionless distance $d_i$ to the ground-truth state vector (Equation~(\ref{eq:euclidean_dist})), \rev{combined in a root mean square over the samples of each impact and averaged} over the twenty impacts ($\bar d$): mean over the three trials for the noisy input and for the denoised output, reduction of the distance (range over the trials), number of impacts improved in every trial, and paired one-tailed $t$-test over the three trial means (2 degrees of freedom). The last column gives the denoised distance of the ablation without the metadata inputs $e$ and $v_{\text{approach}}$.}
\label{tab:impact_distance}
\resizebox{\columnwidth}{!}{%
\begin{tabular}{cccccccc}
\hline
\textbf{Input SNR} & \textbf{Noisy} & \textbf{Denoised} & \textbf{Reduction} & \textbf{Improved} & \textbf{T-Statistic} & \textbf{P-Value} & \textbf{Denoised,} \\
\textbf{(dB)} & $\bar d$ & $\bar d$ & \textbf{(\%)} & \textbf{impacts} & & & \textbf{no metadata} \\ \hline
30.46 & 0.024 & 0.013 & 44--49 & 20/20 & -36.1 & $3.8 \times 10^{-4}$ & 0.018 \\
26.02 & 0.040 & 0.019 & 48--55 & 20/20 & -26.9 & $6.9 \times 10^{-4}$ & 0.024 \\
23.10 & 0.056 & 0.023 & 56--61 & 20/20 & -45.6 & $2.4 \times 10^{-4}$ & 0.030 \\
20.00 & 0.080 & 0.029 & 61--64 & 20/20 & -62.0 & $1.3 \times 10^{-4}$ & 0.039 \\
16.48 & 0.120 & 0.042 & 64--66 & 20/20 & -98.2 & $5.2 \times 10^{-5}$ & 0.056 \\
\hline
\end{tabular}%
}
\end{table}

Table~\ref{tab:impact_physics} resolves the error by state variable and adds a consistency check, in which the contact force is recomputed from the kinematic signals through the Hunt and Crossley law of Equation~(\ref{eq:HC_force}), with the parameters of the simulation, and compared with the clean force; since the law and its parameters are those that generated the data, this is a check of consistency with the generative model, not an independent physical validation. As detailed in Table~\ref{tab:impact_physics}, the results yield the same conclusion: the normalized error falls by 36\% to 73\% for the three state variables and by 47\% to 65\% for the recomputed force, and the fraction of samples that violate the physical constraints is 1.2--2.1\% after denoising against 1.4--4.6\% before, i.e. it is reduced at the noisier levels but not at the least noisy one, where both signals leave about \rev{1.5\%} of the samples slightly below zero near the start and the end of the contact; the denoised force profile provides a more accurate representation of the contact dynamics, effectively recovering the nonlinear impact force profile.

\begin{table}[h!]
\centering
\caption{Per-variable normalized root-mean-square error (NRMSE, i.e. the RMSE divided by the standard deviation of the clean signal), averaged over impacts and trials, for the noisy input and the denoised output: indentation, indentation velocity, measured contact force, and contact force recomputed from the kinematic signals through the Hunt and Crossley law with the parameters of the simulation (consistency with the generative contact law); and fraction of samples that violate $\delta \ge 0$ or $F_N \ge 0$.}
\label{tab:impact_physics}
\resizebox{\columnwidth}{!}{%
\begin{tabular}{cccccc}
\hline
\textbf{Input SNR} & \textbf{NRMSE $\delta$} & \textbf{NRMSE $\dot\delta$} & \textbf{NRMSE $F_N$} & \textbf{NRMSE $F_N$ (law)} & \textbf{Violations (\%)} \\
\textbf{(dB)} & \textbf{noisy / denoised} & \textbf{noisy / denoised} & \textbf{noisy / denoised} & \textbf{noisy / denoised} & \textbf{noisy / denoised} \\ \hline
30.46 & 0.022 / 0.014 & 0.028 / 0.012 & 0.021 / 0.012 & 0.023 / 0.012 & 1.4 / 1.5 \\
26.02 & 0.037 / 0.020 & 0.047 / 0.018 & 0.035 / 0.018 & 0.038 / 0.018 & 2.1 / 1.4 \\
23.10 & 0.052 / 0.025 & 0.065 / 0.023 & 0.049 / 0.022 & 0.054 / 0.022 & 2.7 / 2.1 \\
20.00 & 0.074 / 0.032 & 0.093 / 0.028 & 0.069 / 0.027 & 0.077 / 0.027 & 3.4 / 1.3 \\
16.48 & 0.111 / 0.045 & 0.140 / 0.038 & 0.104 / 0.040 & 0.115 / 0.040 & 4.6 / 1.2 \\
\hline
\end{tabular}%
}
\end{table}

This restoration of signal fidelity is particularly relevant for data-driven discovery tasks. By reducing the variance associated with noise while preserving the underlying nonlinear correlations, the proposed method facilitates the application of algorithms such as Symbolic Regression to rediscover fundamental physical laws from raw measurements, as demonstrated by Lemos \textit{et al.} \cite{lemos2023rediscovering}.

The comprehensive validation across both datasets delimits the operating range of the Time-Window Noise2Noise architecture. On the 3-DOF benchmark the method is effective for input SNRs between 5 and 20 dB, where it improves the signal by 3.1 to 8.9 dB, exceeds the low-pass filter from 15 dB down, and gives the largest gain inside the structural band. At 25 dB input the noise is below the reconstruction floor of the network and the output is worse than the input ($-1.0$ dB), so that, for this architecture, sampling rate, system, and training protocol, the method offers no benefit above about 20 dB; simpler classical filters, or \rev{no denoising} at all, are preferable there. At the lowest level tested, 5 dB, the gain is still 8.9 dB and all three modes remain identifiable after denoising, but under extreme deterioration even substantial SNR gains do not necessarily guarantee physical fidelity, as the signal may become so severely corrupted that stochastic noise dominates deterministic dynamics, violating the temporal coherence assumption underlying self-supervised learning. This lower bound is an observed limitation in these benchmarks, not a general result. Two further requirements follow from the results: the noise must \rev{not be strongly correlated in time}, since \rev{strongly} correlated (brown) noise is reproduced rather than removed, and the record must cover the decay of every mode of interest without a long noise-only tail. Adequate sampling density (Nyquist criterion) and predominantly deterministic governing equations remain necessary; systems with strong stochastic excitation, chaotic dynamics, or irregular sampling may exhibit degraded performance as they violate the smoothness assumptions underlying temporal windowing.

\section{Conclusions}
\label{Conclusions}

This work introduced the Time-Window Noise2Noise (TWN2N) architecture, a blind denoising strategy tailored to dynamic systems. By defining the network input as a temporal sequence rather than an isolated point, and by processing all simultaneously acquired channels with a single network, the model successfully learned to distinguish between the deterministic evolution of physical states and stochastic measurement noise. This was achieved in a fully self-supervised manner, relying solely on the noisy signal itself and on the temporal and multichannel coherence of the measured dynamics, thereby eliminating the constraints of previous methods that required clean ground truth or paired noisy observations.

The main findings, the scope of their validity, and the steps they point to are summarized below.
\begin{itemize}
\item On the linear 3-DOF benchmark, with thirty independent realizations per condition, the gain is positive \rev{in every realization} from 20 dB down to 5 dB input (3.1 to 8.9 dB) and negative at 25 dB ($-1.0$ dB), where the reconstruction floor of the network exceeds the noise; the operating range is therefore bounded above at about 20 dB input SNR for this benchmark and protocol.
\item A zero-phase low-pass filter \rev{with a fixed cutoff at $1.05 f_3$ gains 2.9--3.4 dB between 20 and 5 dB input and 2.0 dB at 25 dB} on a system whose modes span nearly half of the measured bandwidth. TWN2N \rev{matches it at 20 dB input, exceeds it from 15 dB down,} and gives the largest gain inside the structural band (5.7 dB at 15 dB input against \rev{1.1 dB}), lowering the noise floor between the resonances instead of cutting it above the highest mode; the universal wavelet threshold \rev{degrades} these non-sparse responses \rev{from 15 dB input up}.
\item The gain depends on the noise: 2.5 dB under pink noise\rev{, the largest among the methods compared}, and no advantage under brown noise and quantization noise, the former because temporally correlated noise violates the independence assumption on which the blind-spot construction rests, the latter because the rounding error at the level tested is close to the reconstruction floor of the network.
\item For the identification of natural frequencies and damping ratios with the Ibrahim Time Domain method, the frequencies are robust to the noise, whereas the damping ratio of the weakest mode, estimated as $0.084 \pm 0.012$ (mean $\pm$ standard deviation) from the raw signals at 15 dB input against a reference of 0.010, is estimated as $0.0099 \pm 0.0011$ after denoising, with a 25-fold reduction of the dispersion of its frequency; at 20 dB input the denoised estimate is $0.0093 \pm 0.0007$, 7\% below the reference, a bias that the confidence interval of the mean resolves.
\item This recovery holds for the Rayleigh and mass-proportional configurations tested, with damping ratios from 0.002 to 0.050, but not for the stiffness-proportional case, whose fast third mode contributes too few samples above the noise floor of an 8 s record: the record must contain enough samples of the decay of each mode of interest.
\item On the nonlinear impact experiment, the denoised state vector and contact force are closer to the ground truth than the raw measurements at every tested SNR level and in every trial, for a network that knows the test conditions, which indicates that the method preserves the underlying nonlinear correlations while suppressing stochastic interference.
\end{itemize}

The primary contribution of this work is a self-supervised computational method that suppresses noise in signals from deterministic dynamical systems, and thereby improves the damping ratios and contact forces derived from them (the natural frequencies being already robust to the noise on the benchmark), without any clean reference, paired data, or measured noise statistics. Following established practice in blind denoising research \cite{lehtinen2018noise2noise, krull2019noise2void, batson2019noise2self, majumdar2018blind}, the method was validated on controlled synthetic benchmarks where the ground truth is exactly known, which allows the noise regimes to be explored systematically and the preservation of the physical quantities to be quantified. The results are consistent across the linear and the nonlinear system, but they also delimit the method: it helps within an input SNR range bounded above by the reconstruction floor of the network, for noise that is \rev{not strongly correlated} in time, and for records that cover the decay of the modes of interest. The validation on laboratory data is not part of this manuscript.

The window size $p$ and the bottleneck dimension \rev{$\dim(r)$} were fixed throughout from the finite-difference and bottleneck arguments of Section~\ref{Results}, and were deliberately not optimized, since any selection based on the output SNR would require the clean signal that a real measurement does not provide. Future investigations may examine the sensitivity of the method to these two parameters and to the early-stopping rule, provided that the selection criterion is one available without a clean reference, such as the validation loss of the self-supervised training itself. Additionally, alternative architectures incorporating attention mechanisms or adaptive windowing could further enhance the separation of deterministic signals from nonstationary stochastic noise.

\section*{Declarations}

\noindent\textbf{Funding.} This work was supported by the Center for Innovation on New Energies (CINE) and by the Coordena\c{c}\~ao de Aperfei\c{c}oamento de Pessoal de N\'ivel Superior (CAPES), Brazil, Finance Code 001.

\noindent\textbf{Conflict of interest.} The authors declare that they have no conflict of interest.

\noindent\textbf{Ethics approval.} Not applicable.

\noindent\textbf{Consent to participate / Consent for publication.} Not applicable.

\noindent\textbf{Data availability.} The datasets generated and analyzed during the current study are \rev{numerical. They are generated by the scripts of the public repository cited in the Code availability statement, which implement the models described in the manuscript with all their parameters. The values of the individual trials behind the result tables of this work are stored in the same repository, with the script that rebuilds the tables from them.}

\noindent\textbf{Code availability.} \rev{An open-source implementation of the proposed method, together with the scripts that generate the numerical benchmarks and run the principal case studies, is available as a Python library at \url{https://github.com/viniciuserra/Time-Window_Noise2Noise}. The library is archived at Zenodo under the MIT license~\cite{vianna2026twn2ncode}.}

\noindent\textbf{Author contributions.} V.~S.~Vianna: conceptualization, methodology, software, investigation, formal analysis, writing (original draft). T.~H.~Machado: conceptualization, supervision, resources, writing (review and editing). I.~F.~Santos: conceptualization, supervision, funding acquisition, writing (review and editing). All authors read and approved the final manuscript.

\renewcommand\refname{References}
\bibliographystyle{elsarticle-num}
\bibliography{referencias}

\begin{thebibliography}{10}
\expandafter\ifx\csname url\endcsname\relax
  \def\url#1{\texttt{#1}}\fi
\expandafter\ifx\csname urlprefix\endcsname\relax\def\urlprefix{URL }\fi
\expandafter\ifx\csname href\endcsname\relax
  \def\href#1#2{#2} \def\path#1{#1}\fi

\bibitem{lindenbaum1994gabor}
M.~Lindenbaum, M.~Fischer, A.~Bruckstein, On {Gabor's} contribution to image
  enhancement, Pattern Recognition 27~(1) (1994) 1--8.
\newblock \href {https://doi.org/10.1016/0031-3203(94)90013-2}
  {\path{doi:10.1016/0031-3203(94)90013-2}}.

\bibitem{oppenheim2010discrete}
A.~V. Oppenheim, R.~W. Schafer, Discrete-Time Signal Processing, 3rd Edition,
  Pearson, Upper Saddle River, NJ, 2010.

\bibitem{donoho1994ideal}
D.~L. Donoho, I.~M. Johnstone, Ideal spatial adaptation by wavelet shrinkage,
  Biometrika 81~(3) (1994) 425--455.
\newblock \href {https://doi.org/10.1093/biomet/81.3.425}
  {\path{doi:10.1093/biomet/81.3.425}}.

\bibitem{buades2005non}
A.~Buades, B.~Coll, J.-M. Morel, A non-local algorithm for image denoising, in:
  2005 IEEE Computer Society Conference on Computer Vision and Pattern
  Recognition (CVPR'05), Vol.~2, IEEE, 2005, pp. 60--65.
\newblock \href {https://doi.org/10.1109/CVPR.2005.38}
  {\path{doi:10.1109/CVPR.2005.38}}.

\bibitem{savitzky1964smoothing}
A.~Savitzky, M.~J. Golay, Smoothing and differentiation of data by simplified
  least squares procedures, Analytical Chemistry 36~(8) (1964) 1627--1639.
\newblock \href {https://doi.org/10.1021/ac60214a047}
  {\path{doi:10.1021/ac60214a047}}.

\bibitem{fan2020vibration}
G.~Fan, J.~Li, H.~Hao, Vibration signal denoising for structural health
  monitoring by residual convolutional neural networks, Measurement 157 (2020)
  107651.
\newblock \href {https://doi.org/10.1016/j.measurement.2020.107651}
  {\path{doi:10.1016/j.measurement.2020.107651}}.

\bibitem{vincent2008extracting}
P.~Vincent, H.~Larochelle, Y.~Bengio, P.-A. Manzagol, Extracting and composing
  robust features with denoising autoencoders, in: Proceedings of the 25th
  International Conference on Machine Learning, 2008, pp. 1096--1103.
\newblock \href {https://doi.org/10.1145/1390156.1390294}
  {\path{doi:10.1145/1390156.1390294}}.

\bibitem{jain2008natural}
V.~Jain, H.~S. Seung, Natural image denoising with convolutional networks, in:
  D.~Koller, D.~Schuurmans, Y.~Bengio, L.~Bottou (Eds.), Advances in Neural
  Information Processing Systems 21, Curran Associates, Red Hook, NY, 2008, pp.
  769--776.

\bibitem{zhang2017beyond}
K.~Zhang, W.~Zuo, Y.~Chen, D.~Meng, L.~Zhang, Beyond a {Gaussian} denoiser:
  Residual learning of deep {CNN} for image denoising, IEEE Transactions on
  Image Processing 26~(7) (2017) 3142--3155.
\newblock \href {https://doi.org/10.1109/TIP.2017.2662206}
  {\path{doi:10.1109/TIP.2017.2662206}}.

\bibitem{lehtinen2018noise2noise}
J.~Lehtinen, J.~Munkberg, J.~Hasselgren, S.~Laine, T.~Karras, M.~Aittala,
  T.~Aila, {Noise2Noise}: Learning image restoration without clean data, in:
  Proceedings of the 35th International Conference on Machine Learning, Vol.~80
  of Proceedings of Machine Learning Research, PMLR, 2018, pp. 2965--2974.

\bibitem{krull2019noise2void}
A.~Krull, T.-O. Buchholz, F.~Jug, {Noise2Void}---learning denoising from single
  noisy images, in: Proceedings of the IEEE/CVF Conference on Computer Vision
  and Pattern Recognition, 2019, pp. 2129--2137.
\newblock \href {https://doi.org/10.1109/CVPR.2019.00223}
  {\path{doi:10.1109/CVPR.2019.00223}}.

\bibitem{batson2019noise2self}
J.~Batson, L.~Royer, {Noise2Self}: Blind denoising by self-supervision, in:
  International Conference on Machine Learning, PMLR, 2019, pp. 524--533.

\bibitem{vandenende2023self}
M.~van~den Ende, I.~Lior, J.-P. Ampuero, A.~Sladen, A.~Ferrari, C.~Richard, A
  self-supervised deep learning approach for blind denoising and waveform
  coherence enhancement in distributed acoustic sensing data, IEEE Transactions
  on Neural Networks and Learning Systems 34~(7) (2023) 3371--3384.
\newblock \href {https://doi.org/10.1109/TNNLS.2021.3132832}
  {\path{doi:10.1109/TNNLS.2021.3132832}}.

\bibitem{lapins2024dasn2n}
S.~Lapins, A.~Butcher, J.-M. Kendall, T.~S. Hudson, A.~L. Stork, M.~J. Werner,
  J.~Gunning, A.~M. Brisbourne, {DAS-N2N}: machine learning distributed
  acoustic sensing ({DAS}) signal denoising without clean data, Geophysical
  Journal International 236~(2) (2024) 1026--1041.
\newblock \href {https://doi.org/10.1093/gji/ggad460}
  {\path{doi:10.1093/gji/ggad460}}.

\bibitem{li2025blind}
T.~Li, Z.~Liu, Q.~Sui, C.~Lu, J.~Han, S.~Chen, Z.~Li, Deep learning-based blind
  denoising for distributed acoustic sensing seismic data with self-supervised
  and transfer learning, Photonic Sensors 15 (2025).
\newblock \href {https://doi.org/10.1007/s13320-025-0769-x}
  {\path{doi:10.1007/s13320-025-0769-x}}.

\bibitem{wang2024seismic}
D.~Wang, G.~Chen, J.~Chen, Q.~Cheng, Seismic data denoising using a
  self-supervised deep learning network, Mathematical Geosciences 56 (2024)
  487--510.
\newblock \href {https://doi.org/10.1007/s11004-023-10089-3}
  {\path{doi:10.1007/s11004-023-10089-3}}.

\bibitem{langarica2023contrastive}
S.~Langarica, F.~N{\'u}{\~n}ez, Contrastive blind denoising autoencoder for
  real time denoising of industrial {IoT} sensor data, Engineering Applications
  of Artificial Intelligence 120 (2023) 105838.
\newblock \href {https://doi.org/10.1016/j.engappai.2023.105838}
  {\path{doi:10.1016/j.engappai.2023.105838}}.

\bibitem{yang2024blind}
D.~Yang, H.~R. Karimi, D.~Ma, A blind denoising method for noisy rotating
  machinery vibration signals, IFAC-PapersOnLine 58~(4) (2024) 740--745.
\newblock \href {https://doi.org/10.1016/j.ifacol.2024.07.308}
  {\path{doi:10.1016/j.ifacol.2024.07.308}}.

\bibitem{goodfellow2016deep}
I.~Goodfellow, Y.~Bengio, A.~Courville, Deep Learning, MIT Press, Cambridge,
  MA, 2016.

\bibitem{majumdar2018blind}
A.~Majumdar, Blind denoising autoencoder, IEEE Transactions on Neural Networks
  and Learning Systems 30~(1) (2019) 312--317.
\newblock \href {https://doi.org/10.1109/TNNLS.2018.2838679}
  {\path{doi:10.1109/TNNLS.2018.2838679}}.

\bibitem{chiang2019noise}
H.-T. Chiang, Y.-Y. Hsieh, S.-W. Fu, K.-H. Hung, Y.~Tsao, S.-Y. Chien, Noise
  reduction in {ECG} signals using fully convolutional denoising autoencoders,
  IEEE Access 7 (2019) 60806--60813.
\newblock \href {https://doi.org/10.1109/ACCESS.2019.2912036}
  {\path{doi:10.1109/ACCESS.2019.2912036}}.

\bibitem{ashfahani2020devdan}
A.~Ashfahani, M.~Pratama, E.~Lughofer, Y.-S. Ong, {DEVDAN}: Deep evolving
  denoising autoencoder, Neurocomputing 390 (2020) 297--314.
\newblock \href {https://doi.org/10.1016/j.neucom.2019.07.106}
  {\path{doi:10.1016/j.neucom.2019.07.106}}.

\bibitem{cho2013simple}
K.~Cho, Simple sparsification improves sparse denoising autoencoders in
  denoising highly corrupted images, in: International Conference on Machine
  Learning, PMLR, 2013, pp. 432--440.

\bibitem{vianna2026twn2ncode}
V.~S. Vianna, T.~H. Machado, I.~F. Santos, {TWN2N}: {T}ime-{W}indow
  {N}oise2{N}oise, version 1.0.0, Zenodo,
  \url{https://doi.org/10.5281/zenodo.24355387} (2026).

\bibitem{hunt1975coefficient}
K.~H. Hunt, F.~R.~E. Crossley, Coefficient of restitution interpreted as
  damping in vibroimpact, Journal of Applied Mechanics 42~(2) (1975) 440--445.
\newblock \href {https://doi.org/10.1115/1.3423596}
  {\path{doi:10.1115/1.3423596}}.

\bibitem{zhang2009validation}
Y.~Zhang, I.~Sharf, Validation of nonlinear viscoelastic contact force models
  for low speed impact, Journal of Applied Mechanics 76~(5) (2009) 051002.
\newblock \href {https://doi.org/10.1115/1.3112739}
  {\path{doi:10.1115/1.3112739}}.

\bibitem{jcgm2008gum}
{JCGM}, Evaluation of measurement data --- {G}uide to the expression of
  uncertainty in measurement ({GUM}), JCGM 100:2008, {J}oint Committee for
  Guides in Metrology, BIPM (2008).

\bibitem{ibrahim1977method}
S.~R. Ibrahim, E.~Mikulcik, A method for the direct identification of vibration
  parameters from the free response, The Shock and Vibration Bulletin 47~(4)
  (1977) 183--198.

\bibitem{ewins2009modal}
D.~J. Ewins, Modal testing: theory, practice and application, John Wiley \&
  Sons, 2009.

\bibitem{donoho1995noising}
D.~L. Donoho, De-noising by soft-thresholding, IEEE Transactions on Information
  Theory 41~(3) (1995) 613--627.
\newblock \href {https://doi.org/10.1109/18.382009}
  {\path{doi:10.1109/18.382009}}.

\bibitem{daubechies1992ten}
I.~Daubechies, Ten lectures on wavelets, SIAM, 1992.
\newblock \href {https://doi.org/10.1137/1.9781611970104}
  {\path{doi:10.1137/1.9781611970104}}.

\bibitem{lemos2023rediscovering}
P.~Lemos, N.~Jeffrey, M.~Cranmer, S.~Ho, P.~Battaglia, Rediscovering orbital
  mechanics with machine learning, Machine Learning: Science and Technology
  4~(4) (2023) 045002.
\newblock \href {https://doi.org/10.1088/2632-2153/acfa63}
  {\path{doi:10.1088/2632-2153/acfa63}}.

\end{thebibliography}
\newpage
\appendix

\subsection*{Appendix A}
\section*{Neural network architecture and training details}
\addcontentsline{toc}{section}{Appendix A: Deep learning model architecture and hyperparameter settings}
\label{appendix:architecture}

This appendix specifies the network architectures and training procedures for both the 3-DOF mass-spring-damper system and the impact experiment dataset. The models were implemented using the Keras framework with a TensorFlow backend.

\subsection*{3-DOF system models}
For the 3-DOF dataset, two distinct architectures were evaluated: a standard baseline autoencoder and the proposed Time-Window Noise2Noise model. 

The baseline standard autoencoder processes a 6-dimensional input vector (positions and accelerations for 3 masses) through a 4-dimensional latent space. The Time-Window Noise2Noise model uses $p=2$, yielding a 24-dimensional input vector (2 state variables $\times$ 3 masses $\times$ 4 neighboring time steps). This input is also mapped to a 4-dimensional latent space before being reconstructed into a single time-step output (6 dimensions).

Regarding the nonlinear transformations, all hidden layers in both the encoder and decoder stages utilized the hyperbolic tangent ('tanh') activation function. The output layer employed a linear activation function to allow for the reconstruction of the continuous dynamic signals without range constraints. The specific layer-wise configuration for both models is presented in Table \ref{tab:arch_3dof}.

\begin{table}[h!]
\centering
\caption{Detailed architecture of the Dense layers for the 3-DOF system models.}
\label{tab:arch_3dof}
\resizebox{\columnwidth}{!}{%
\begin{tabular}{lccc}
\hline
\textbf{Layer Type} & \textbf{Baseline Autoencoder} & \textbf{Time-Window Noise2Noise} & \textbf{Activation} \\ 
 & \textit{Input Dim: 6 / Output Dim: 6} & \textit{Input Dim: 24 / Output Dim: 6} & \\ \hline
Input Layer & 6 & 24 & - \\
Dense (Encoder) & 114 & 102 & tanh \\
Dense (Encoder) & 54 & 54 & tanh \\
Dense (Encoder) & 54 & 54 & tanh \\
Dense (Encoder) & 8 & 12 & tanh \\
\textbf{Latent Space} & \textbf{4} & \textbf{4} & \textbf{tanh} \\
Dense (Decoder) & 8 & 12 & tanh \\
Dense (Decoder) & 24 & 30 & tanh \\
Dense (Decoder) & 54 & 54 & tanh \\
Dense (Decoder) & 54 & 54 & tanh \\
Output Layer & 6 & 6 & linear \\ \hline
Total Parameters & 15,360 & 17,218 & - \\ \hline
\end{tabular}%
}
\end{table}

\subsection*{Impact experiment model}
For the nonlinear impact dataset (Hunt-Crossley model), the network was similarly designed with \texttt{tanh} activation functions for all hidden layers to capture the nonlinearities of the contact force, while the output layer utilized a linear activation. The input dimension was $N_{in}=20$, and the latent dimension was set to $\dim (r)=3$. The architecture follows a symmetric expansion-compression logic relative to the input size, as detailed in Table \ref{tab:arch_contact}.

\begin{table}[h!]
\centering
\caption{Time-Window Noise2Noise configuration for the Impact Experiment dataset.}
\label{tab:arch_contact}
\resizebox{0.53\columnwidth}{!}{%
\begin{tabular}{lcc}
\hline
\textbf{Layer Type} & \textbf{Layer Size (Neurons)} & \textbf{Activation} \\ \hline
Input Layer & 20 & - \\
Dense (Encoder) & 45 & tanh \\
Dense (Encoder) & 25 & tanh \\
Dense (Encoder) & 25 & tanh \\
Dense (Encoder) & 3 & tanh \\
\textbf{Latent Space} & \textbf{3} & \textbf{tanh} \\
Dense (Decoder) & 3 & tanh \\
Dense (Decoder) & 25 & tanh \\
Dense (Decoder) & 25 & tanh \\
Dense (Decoder) & 45 & tanh \\
Output Layer & 5 & linear \\ \hline
Total Parameters & 4,997 & - \\ \hline
\end{tabular}%
}
\end{table}

\subsection*{Training process and hyperparameters}

The optimization was performed using the Adam algorithm with a learning rate of $\alpha = 0.001$ and a mini-batch size of 32 samples. The Mean Absolute Error (MAE) was selected as the loss function to minimize the reconstruction error. The maximum number of epochs was set to 5000.

\subsubsection*{Custom early-stopping strategy}
To prevent the model from memorizing the stochastic noise (overfitting) while ensuring sufficient learning of the underlying dynamics, a custom callback named \texttt{StopAfterOverfitting} was implemented. Unlike standard early stopping, this strategy requires two conditions for termination:

\begin{enumerate}
    \item \textbf{Stagnation:} The validation loss has not improved for a defined number of epochs (patience).
    \item \textbf{Divergence:} The validation loss is strictly greater than the training loss ($Loss_{val} > Loss_{train}$).
\end{enumerate}

The training halts only when both conditions are met, ensuring that the stopping point corresponds to a regime where generalization performance begins to degrade relative to training performance. Upon termination, the weights from the epoch with the lowest validation loss are restored.

The patience parameter was set to 100 epochs for both datasets. This strategy proved essential for obtaining the denoising results presented in the main text, acting as an implicit regularization mechanism. With this rule, the 450 trainings of the 3-DOF study stopped after a median of 270 epochs (range 124 to 905), the epoch with the lowest validation loss being the 170th in median (range 24 to 805), and each training took between 0.6 and 10 minutes on two CPU cores (median 2.4 minutes); the stopping point therefore adapts to each record without any access to the clean signal. For the impact dataset (299,522 samples), the 15 trainings stopped after a median of 245 epochs (range 154 to 395) and took between 35 and 90 minutes each on two CPU cores (median 58 minutes).

\end{document}